\documentclass[12pt]{article}
\usepackage{graphics, color}
\usepackage{amssymb}

\def\Xint#1{\mathchoice {\XXint\displaystyle\textstyle{#1}}%
{\XXint\textstyle\scriptstyle{#1}}%
{\XXint\scriptstyle\scriptscriptstyle{#1}}%
{\XXint\scriptscriptstyle\scriptscriptstyle{#1}}%
\!\int} 
\def\XXint#1#2#3{{\setbox0=\hbox{$#1{#2#3}{\int}$} 
\vcenter{\hbox{$#2#3$}}\kern-.5\wd0}} 
 
\def\dashint{\Xint-} 

\newcommand{\sect}[1]{\section{#1}\setcounter{equation}{0}}

\newcommand{\OL}[1]{ \hspace{.5pt}\overline{\hspace{-.5pt}#1
     \hspace{-.5pt}}\hspace{.5pt} }

\input epsf

\begin{document}


\begin{titlepage}
\bigskip
\rightline{}
\rightline{hep-th/0508232}
\bigskip\bigskip\bigskip
\centerline{\Large \bf Bethe Ansatz for}
\medskip
\centerline{\Large \bf a Quantum Supercoset Sigma Model}
\bigskip\bigskip\bigskip

\centerline{\large Nelia Mann$^a$ and Joseph Polchinski$^b$}
\bigskip
\centerline{\em $^a$ Department of Physics, University of California,} \centerline{\em Santa Barbara, CA 93106} \centerline{\em nelia@physics.ucsb.edu}
\smallskip
\centerline{\em $^b$ Kavli Institute for Theoretical Physics, University of California,} 
\centerline{\em Santa Barbara, CA 93106} \centerline{\em joep@kitp.ucsb.edu}
\bigskip
\bigskip
\bigskip\bigskip


\begin{abstract}
We study an integrable conformal ${OSp}(2m + 2|2m)$ supercoset model as an analog to the $AdS_5 \times S^5$ superstring world-sheet theory.  Using the known S-matrix for this system, we obtain integral equations for states of large particle number in an $SU(2)$ sector, which are exact in the sigma model coupling constant.  As a check, we derive as a limit the general classical Bethe equation of Kazakov, Marshakov, Minahan, and Zarembo.  There are two distinct quantum expansions around the well-studied classical limit, the $\lambda^{-1/2}$ effects and the $1/J$ effects.  Our approach captures the first type, but not the second.

\medskip
\noindent
\end{abstract}
\end{titlepage}
\baselineskip=16pt

\sect{Introduction}

The discovery of integrable structures in both the gauge theory~\cite{MZ,BKS,BS1} and string theory~\cite{MSW,BPR} limits of the AdS/CFT duality gives a strong hint that ${\cal N} = 4$ supersymmetric gauge theory is solvable, at least in the planar approximation.\footnote{For earlier work on integrability in QCD see refs.~\cite{QCD} and the review~\cite{QCDrev}.}  Subsequently this subject has advanced on many fronts; for reviews see refs.~\cite{arkrev,Mrev,Brev,Trev,Zrev,Srev,Prev,Bstrings}.

Let us note here a few key developments, particularly those concerning the string side and its relation to the gauge side.  For states of large charge, it has been possible to compare the operator dimensions obtained in the gauge and string descriptions~\cite{BMN,GKP,FT,FT2,BMSZ,BFST}.  The classical string picture can be derived directly by going to a coherent state representation for the gauge theory operators~\cite{Kru}.
However, higher order calculations show that the gauge-string correspondence is not a simple as initially assumed~\cite{Cetal,SS}.  In the string sigma model, the nonlocal conserved charges can be used to construct a spectral curve that characterizes the general classical solution~\cite{KMMZ,BKSZ,BKSZ2}.  The Bethe ansatz equation for this spectral curve has been further developed and compared with the gauge theory Bethe ansatz~\cite{BDipS,SSmat,BS05}.  It has been argued that the nonlocal charges are conserved in the full quantum sigma model for the $AdS_5 \times S^5$ string~\cite{Vallilo,Berk}.  In refs.~\cite{AFS,Bqu} an extension of the Bethe ansatz to the quantized sigma model is conjectured, but the discrepancy with the gauge theory remains.  Recently there has been further study of the one-loop quantum corrections to the sigma model, again with apparent discrepancies~\cite{oneloop}.
Finally, additional discussions of the sigma model conserved charges and their relation to the gauge theory charges can be found in refs.~\cite{charges}.

To summarize, integrability is fairly well developed for the classical sigma model, but the extension to the quantum sigma model is in very preliminary state.  There has been a focus on quantities for which the quantum corrections are hoped to take a rather restricted form~\cite{FT}, but ultimately it is clear that most of the physics of the AdS/CFT system is dependent on the quantization of the sigma model.  Thus in this paper we wish to take a complementary approach, starting with a sigma model where some integrable structure is already known at the quantum level.  This is the $OSp(2m+2|2m)$ coset model~\cite{MP}, specifically $OSp(2m+2|2m)/OSp(2m+1|2m)$, whose bosonic part is $S^{2m+1}$.  Like the $AdS_5 \times S^5$ world-sheet theory it is conformally invariant and its target space is a supergroup coset.  It is a different coset, and lacks the ghost and BRST structure of the string theory, but still is likely to give a hint of the structure that will appear in the full string theory.

In the coset model the integrable structure takes the form of an S-matrix.\footnote{The importance of the world-sheet S-matrix has recently been emphasized in ref.~\cite{SSmat}.}  This is derived by taking the conformal $n \to 2$ limit of the $OSp(2m+n|2m)$ S-matrix.  The latter~\cite{SWK} is obtained from the well-known $O(n)$ S-matrix~\cite{ZZS} by addition of equal numbers of bosonic and fermionic coordinates.  In ref.~\cite{MP} it was shown that the $n \to 2$ limit can be defined, and the resulting S-matrix used in a finite-density Bethe ansatz.  The limit has the feature that, in addition to the right-moving and left-moving particles that would be expected in a conformal theory, there is a continuum of zero-energy states, so-called `zero modes'~\cite{MP} though perhaps `non-movers' would be more apt.  

Ref.~\cite{MP} considered only a $U(1)$ sector of the sigma model, which is trivial from the point of view of the analogous string theory.  In this paper we extend the analysis to an $SU(2)$ sector.  We obtain the Bethe ansatz equations for the full quantum sigma model, and then take the classical limit.  The classical theory, a bosonic sigma model on an $S^3$, is identical to the 
$SU(2)$ sector of the $AdS_5 \times S^5$ theory.\footnote{To obtain nontrivial physical states we imagine appending a free timelike coordinate.  We could also analytically continue in the charges, equivalent to spinning strings on $AdS_3$.}
  Indeed, we recover the Bethe equation found in refs.~\cite{BKSZ}.  The embedding of this Bethe ansatz into a quantum theory is our main result.  

One important lesson is that the extension of the classical Bethe equations to the quantum theory involves two separate deformations.  The classical sigma model here is the classical {\it field} limit, in which the coupling is taken to zero and the number of quanta is taken to infinity.  Thus to recover the quantum theory we must restore finite quantum numbers (that is, $1/J$ corrections), and also include world-sheet quantum effects ($g^2$ corrections).\footnote{To see that the latter are independent effects, consider the world-sheet theory on a line rather than a circle.  At finite density the $1/J$ corrections are strictly absent --- the Bethe ansatz remains continuous --- but the physics certainly depends on $g^2$.  Also, the three loop discrepancy~\cite{Cetal,SS} is visible in the continuous Bethe equations~\cite{BDipS}, and so should be due to $g^2$ effects.}  Refs.~\cite{AFS,Bqu} focussed on the $1/J$ corrections.  We are unable, in our current work, to address these, but we have a full account of the $g^2$ corrections.

In section~2 we review the $OSp(2m+2|2m)$ coset model, the use of the S-matrix, and the Bethe ansatz in the $U(1)$ sector.  Most of the results are from ref.~\cite{MP}, though we treat the classical limit in more detail.  In section~3 we obtain the Bethe ansatz for the $SU(2)$ sector, as well as its reductions to single impurities and to nonrelativistic impurities.  In section~4 we develop the classical limit.  The zero modes enter in an interesting way: the somewhat complicated form for the Bethe equation given in ref.~\cite{BDipS} arise from a simpler equation when they are integrated out.  In the appendix we review the finite Hilbert transform, which appears in the classical limit.

One might wonder whether the agreement between the $SU(2)$ sectors of our model and the $AdS_5 \times S^5$ theory, which must hold at the classical level, might fortuitously extend to the quantum theory.  In fact this is unlikely.  In our model there is no supersymmetry connecting the spacelike $S^3$ and the appended time coordinate, so the quantum corrections should take a less restricted form.  In addition, our model appears to have a phase transition at finite world-sheet coupling~\cite{MP}, as we will discuss further in section~2.  Such a transition is not expected in the string theory.  We are currently attempting to extend our approach to the $PSL(m|m)$ model~\cite{BVW,BSZ,zirn}, whose symmetry structure is closer to the string theory.  The Bethe ansatz takes a somewhat different form, and the phase transition may be absent.

\sect{Overview and review}

\subsection{The supercoset model}

Consider a nonlinear sigma model based on a field $\varphi_i$ whose first $2m+n$ components are commuting and whose last $2m$ components are anticommuting.
The action and constraint are
\begin{equation}
S = -\frac{1}{2g^2} \int d\tau\, d\sigma\, J^{ij} \partial_\mu \varphi_i \partial^\mu
\varphi_j \ ,\quad  J^{ij} \varphi_i\varphi_j = 1\ ,
\end{equation}
where
\begin{equation}
J^{ij} = \left[
\begin{array}{ccc}
I_{2m+n}  &  0 & 0  \\
 0 & 0  & -I_m  \\
0  & I_m & 0  
\end{array}
\right]\ .
\end{equation}
The action is invariant under an $OSp(2m+n|m)$ symmetry.  Correlation functions of fields restricted to a subset of $n$ bosonic components are identical to those of the bosonic $O(n)$ coset model, because the path integral over the remaining $2m$ bosonic coordinates is the reciprocal of the integral over the fermionic coordinates~\cite{ParSour,Weg}.
In particular, the $OSp(2m+2|m)$ is conformally invariant, because the $O(2)$ model is free.  
However, it is not rational: it is conformally invariant without a Wess-Zumino term, and the separate right- and left-moving currents are not conserved. Instead it possesses an infinite family of nonlocal charges constructed from a flat connection, by direct generalization of the construction for the $O(n)$ coset~\cite{LPohl,BIZZ}.

For quantum sigma models, the integrable structure is encoded in a factorizable S-matrix~\cite{ZZS}.  For massless theories, the usual definition of the S-matrix does not apply because particles moving in the same direction do not separate.  Nevertheless, the massless limit of the S-matrix of a massive integrable theory can still be used in the finite density Bethe ansatz~\cite{ZZ1,FS}: it retains its interpretation as the relative phase acquired in the wavefunction when one particle is moved past another.  

The flat spacetime S-matrix does not directly give the full set of amplitudes needed on the string world-sheet because the string has finite spatial volume while the S-matrix is defined in infinite volume.  In a relativistic field theory the vacuum is nontrivial, and so in finite volume the virtual particle states shift; one signature of this is the Casimir energy.  There do not yet exist general methods to account for this shift and construct the finite volume system.  Thus the questions that are readily answered involve states with a large number $K$ of real particles, where the effect of the virtual particles represent a relative fraction $1/K$.  It is not necessary that there be a large net charge $J$, but in fact the Bethe ansatz is simplest when all particles have the same sign of the charge, and so we will focus on this case.  Thus in our present work we are insensitive to $1/J$ corrections; going beyond this is an important future direction.

\begin{figure}
\begin{center}
\resizebox{5.5in}{!}{\epsfbox{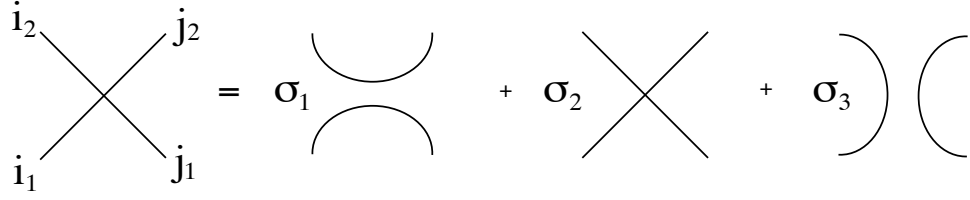}}
\end{center}
\caption{\label{S-matrix} Terms in the S-matrix}
\end{figure}

The exact S-matrix for the $O(n)$ model is well-known~\cite{ZZS}, and the ${OSp}(2m + n|2m)$ symmetry allows this to be lifted in a unique way to the supercoset model~\cite{SWK}.\footnote{The supercoset theory is only pseudounitary, because the indefinite metric $J^{ij}$ appears and we have no analog of $\kappa$ symmetry to remove the unwanted states, but the S-matrix is still defined, and factorizable.}
The S-matrix has three terms,
\begin{eqnarray}
| i_1\,\theta, j_1\, \theta'; {\rm in} \rangle &=& S^{j_2 i_2}_{j_1 i_1}(\theta - \theta') 
| j_2\,\theta, i_2\, \theta'; {\rm out} \rangle\ ,
\nonumber\\[1mm]
S(\theta) &=& \sigma_1(\theta) E + \sigma_2(\theta) P + \sigma_3(\theta) I
\label{eq:ospS}
\end{eqnarray}
where
\begin{eqnarray}
E^{j_2 i_2}_{j_1 i_1} & = & J_{i_1 j_1} J^{i_2 j_2} \\
P^{j_2 i_2}_{j_1 i_1} & = & \delta^{j_2}_{i_1}\delta^{i_2}_{j_1} \\
I^{j_2 i_2}_{j_1 i_1} & = & (-1)^{p_{i_1}+ p_{j_1}}
\delta^{i_2}_{i_1}\delta^{j_2}_{j_1}\ ;
\end{eqnarray}
here $p_i$ is 0 for a bosonic component and 1 for a fermionic component.  The tensor structures are shown diagrammatically in fig.~\ref{S-matrix}.
The functions $\sigma_{i}(\theta)$ are
\begin{eqnarray}
\sigma_1 &=& -\frac{2i\pi}{(n-2)(i\pi - \theta)}\sigma_2\ ,
\quad
\sigma_3 = -\frac{2i\pi}{(n - 2)\theta}\sigma_2\ ,
\nonumber\\[3pt]
\sigma_2 &=& \frac{\Gamma\left(1 - \frac{\theta}{2i\pi}\right)\Gamma\left(\frac{1}{2} + \frac{\theta}{2i\pi}\right)\Gamma\left(\frac{1}{n - 2} + \frac{\theta}{2i\pi}\right)\Gamma\left(\frac{1}{2} + \frac{1}{n - 2} - \frac{\theta}{2i\pi}\right)}{\Gamma\left(\frac{\theta}{2i\pi}\right)\Gamma\left(\frac{1}{2} - \frac{\theta}{2i\pi}\right)\Gamma\left(1 + \frac{1}{n - 2} - \frac{\theta}{2i\pi}\right)\Gamma\left(\frac{1}{2} + \frac{1}{n - 2} + \frac{\theta}{2i\pi}\right)}\ . \label{eq:sigma2}
\end{eqnarray}

The parameter $n$ in the S-matrix can be treated as a continuous parameter in the Bethe ansatz and in Feynman diagrams, with the definition that Supertrace(1)$\equiv \sum_i (-1)^{p_i} = n$.  In particular the Yang-Baxter equation is satisfied.  Since the $n=2$ theory is conformal, the $\beta$ function is of the form $\beta(g) = (n-2)b(g)$ where $b(g)$ is finite as $n \to 2$.  
The coupling thus runs arbitrarily slowly as $n \to 2$: it is a function of 
\begin{equation}
\chi = (n-2) \ln(E/M)
\end{equation}
where $M$ is the dynamically generated mass.  For example, from the one-loop beta function it follows that
\begin{equation}
g^2 = \frac{2\pi}{\chi} + O\Biggl(\frac{\ln\chi}{\chi^2}\Biggr) \label{eq:onel}
\end{equation}
at large $\chi$.
By holding $\chi$ and $E$ fixed as $n \to 2$ and $M \to 0$, we obtain a limit in which the coupling takes the constant value $g(\chi)$.  In particular we get the same coupling if we use another reference energy $E'$ where $E'/E$ is fixed, since $(n-2) \ln(E'/E)$ goes to zero.

In particular, holding fixed the single-particle energy $\varepsilon = M \cosh \theta$ implies that we hold fixed one of 
\begin{eqnarray}
\tilde\theta_R &=& \theta - \chi/(n-2)\ ,\quad \varepsilon = \frac{\mu}{2} e^{\tilde\theta_R}
\ ,\nonumber\\
\tilde\theta_L &=& \theta + \chi/(n-2)\ ,\quad \varepsilon = \frac{\mu}{2} e^{-\tilde\theta_L}
\  .
\end{eqnarray}
Thus the excitations that carry energy and momentum separate into a right-moving range with fixed $\tilde\theta_R$ and a left-moving range with fixed $\tilde\theta_L$.  The rapidity difference between these two sets diverges as $1/(n-2)$.  The surprising result in ref.~\cite{MP}, which we will review below, is that in the limit there remains also a {\it continuum} of `zero-mode' excitations between the right- and left-movers.  For these, $\phi = (n-2)\theta$ is held finite, where $-\chi < \phi < \chi$.  The zero modes do not carry single-particle energies, but they affect the total energy through their interaction with the right- and left-movers.  We denote the three types of particle state by $R$, $L$, and $0$.

\subsection{The $U(1)$ sector}

We begin by building states with a finite density of excitations all positively charged under a single $U(1) = O(2) \subset O(2m+n)$, for example states created by $\varphi_1 + i \varphi_2$.  Acting on these, $P=I$ and $E=0$, giving
\begin{equation}
S_{pp}(\theta)  = \sigma_2(\theta) + \sigma_3(\theta) = \frac{\Gamma\left(\frac{1}{n - 2} - \frac{i\theta}{2\pi}\right)\Gamma\left(\frac{1}{n - 2} + \frac{1}{2} + \frac{i\theta}{2\pi}\right)\Gamma\left(\frac{1}{2} - \frac{i\theta}{2\pi}\right)\Gamma\left(\frac{i\theta}{2\pi}\right)}{\Gamma\left(\frac{1}{n - 2} + \frac{i\theta}{2\pi}\right)\Gamma\left(\frac{1}{n - 2} + \frac{1}{2} - \frac{i\theta}{2\pi}\right)\Gamma\left(\frac{1}{2} + \frac{i\theta}{2\pi}\right)\Gamma\left(\frac{-i\theta}{2\pi}\right)}\ . \label{eq:so2}
\end{equation}

The standard Bethe ansatz equation for a state of identical particles, from periodicity on a space of length $L$, is \cite{bethe, thacker}
\begin{equation} \label{eq: beginning}
e^{ip_{j}L} =\prod_{i \ne j} S(\theta_{i} - \theta_{j})\ . 
\end{equation}
Here $\theta_{j}$ is the rapidity of the $j$th particle and $p_{j} = m\sinh \theta_{j}$.  Following standard steps we take the logarithm of eq.~(\ref{eq: beginning}),
\begin{equation} 
p_{j}L = -i \sum_{i \ne j} \ln S(\theta_{i} - \theta_{j}) + 2\pi l_j\ . \label{logbet}
\end{equation}
Each rapidity $\theta_{j}$ is thus associated with an integer $l_{j}$ from the branch cut in the logarithm.  For $-i \ln S$ we fix the branch that increases monotonically from $0$ to $2\pi$.  
In the present discussion we focus on a single filled band particle states.  In this case the integers are consecutive, $l_{j+1} = l_{j} + 1$, and the rapidities are found to increase monotonically with $j$~\cite{yy}.  

For future reference we can also write this in another way.  Suppose that we define the logarithm differently, so that it increases from $0$ to $\pi$ (taking the latter value at $\theta =  0$), then jumps to $-\pi$ and finally increases to $0$,
\begin{equation}
-i\, \hat{\ln} S(\theta) = - i
\,{\ln} S(\theta) - 2\pi \Theta(\theta)\ . \label{eq:hatln}
\end{equation}
The integer $l_j$ now takes a constant value $\hat l$ for all particles, as the jump by $2\pi$ on the right-hand side of eq.~(\ref{logbet}) as we move from $j$ to $j+1$ now comes from $ -i\ln S(\theta_{j+1} - \theta_{j})$.  We will denote the logarithm with this definition by $\hat{ \ln}$.  For particle distributions consisting of several filled bands, $\hat l$ is a different constant for each band.

Now take the thermodynamic limit of a large number of particles, holding the density fixed and letting $L\to\infty$.  The difference between consecutive rapidities becomes small with the density (per rapidity and length) finite,
\begin{equation}
\rho(\theta_{j}) = \frac{1}{L(\theta_{j+1} - \theta_{j})}\ .
\end{equation}
These finite density states obey
\begin{equation} \label{eq: continuous}
\frac{M}{2\pi}\cosh \theta = \rho(\theta) + \int_{-B_{L}}^{B_{R}} K(\theta - \theta') \rho(\theta') \ d\theta'\ ,
\end{equation}
where
\begin{equation}
K(\theta) = \frac{1}{2\pi i} \frac{d}{d\theta} \ln S(\theta)\ . \label{eq:kdef}
\end{equation}
This is valid only in the range $-B_{L} < \theta < B_{R}$ where the particle states are filled; outside this range $\rho(\theta) = 0$.

We now take the $n\to2$ limit; for more details see ref.~\cite{MP}.  From the discussion in section~2.1, we see that there are are some rapidity differences that remain fixed in the limit, and others that scale as $1/(n-2)$.  In fact, both limits of $S_{pp}$ are nontrivial, because some of the gamma functions~(\ref{eq:so2}) contain $1/(n-2)$ in their argument and others do not.  Specifically,
\begin{equation}
S_{\rm I}(\theta) \equiv \lim_{n \to 2} S_{pp}(\theta) = \frac{\Gamma\left(\frac{1}{2} - \frac{i\theta}{2\pi}\right)\Gamma\left(\frac{i\theta}{2\pi}\right)}{\Gamma\left(\frac{1}{2} + \frac{i\theta}{2\pi}\right)\Gamma\left(\frac{-i\theta}{2\pi}\right)}\ .
\end{equation}
and
\begin{equation}
S_{\rm II}(\phi) \equiv \lim_{n \to 2} S_{pp}(\phi/[n-2])
 = \left(\frac{2\pi + i\phi}{2\pi - i\phi}\right)^{1/2}e^{i\pi \ \mbox{\scriptsize{sign}} (\phi)/2}\ .
\end{equation}
In particular, $S_{\rm I}$ appears in $RR$ and $LL$ scattering, $S_{RR} = S_{LL} = S_{\rm I}$, while $S_{\rm II}$ appears in $RL$, $R0$, $00$, and $0L$ scattering.

The Bethe ansatz equation separates into integral equations for right-movers, left-movers, and zero-modes:\footnote{The existence of these particles in the middle is implied by the fact that the limit of $S_{RR}$ as the rapidity difference them large is not equal to $S_{RL}$.  Thus some states must get ``trapped'' in between in the conformal limit.}  
\begin{eqnarray} 
\rho_{R}(\theta) -  \int_{-\infty}^{\tilde{B}_{R}} K_{RR}(\theta - \theta')\rho_{R}(\theta') \, d\theta'
&=& \frac{\mu}{4\pi}e^{\theta} \ ,\nonumber\\
\rho_{L}(\theta) -  \int_{-\tilde{B}_{L}}^{\infty} K_{LL}(\theta - \theta') \rho_{L}(\theta') \ , d\theta&=& \frac{\mu}{4\pi}e^{-\theta} \label{eq:rightleft}
\end{eqnarray}
and
\begin{equation} \label{eq:zmeq}
\frac{1}{2}\rho_{0}(\phi) - \int_{-\chi}^{\chi} \frac{\rho_{0}(\phi')}{4\pi^2 + (\phi - \phi')^2} \ d\phi'= \frac{{\cal J}_R}{4\pi^2 + (\phi - \chi)^2} + \frac{{\cal J}_L}{4\pi^2 + (\phi + \chi)^2} \ .
\end{equation}
We have defined the fixed endpoints $\tilde{B}_R = B_R - \chi/(N-2)$,  $\tilde{B}_L = B_L - \chi/(N-2)$, and the number densities
 \begin{equation}
{\cal J}_{R} = \int_{-\infty}^{\tilde{B}_R} \rho_{R}(\theta) \, d\theta\ , \ \ \ {\cal J}_{L} = \int_{-\tilde{B}_{L}}^{\infty} \rho_{L}(\theta) \, d\theta\ ,\ \ \  {\cal J}_{0} = \int_{-\chi}^{\chi} \rho_{0}(\phi) \ d\phi\ .
\end{equation}

The semi-infinite integral equations for the right- and left-movers can be solved by the Wiener-Hopf method.  This is done in ref.~\cite{MP}; for the present paper we need only the result for the energy and momentum densities in terms of the number densities:
\begin{equation} \label{eq: PandE}
 \mathcal{P} \equiv P/L = \frac{\pi({\cal J}_R^2 - {\cal J}_L^2)}{2}\ , \quad
\mathcal{E} \equiv E/L = \frac{\pi({\cal J}_{R}^2 + {\cal J}_{L}^2)}{2}\ ,
\end{equation}
so that
\begin{equation}
\mathcal{E}
=  \frac{\pi ({\cal J}_{R} + {\cal J}_{L})^2}{4} + \frac{\mathcal{P}^2}{\pi({\cal J}_{R} + {\cal J}_{L})^2} \ . \label{eq:energ}
\end{equation}  
The zero mode equation determines ${\cal J}_{0}$ in the form 
\begin{equation}
{\cal J}_{0} = (h(\chi) - 1)({\cal J}_{L} + {\cal J}_{R})
\end{equation}
for some function $h(\chi)$.  This in turn gives ${\cal J}_{L} + {\cal J}_{R} = {\cal J}/h(\chi)$, and so
\begin{equation}
\mathcal{E}
=  \frac{\pi {\cal J}^2}{4 h(\chi)^2} + \frac{\mathcal{P}^2  h(\chi)^2}{\pi {\cal J}^2} 
\label{eq:epj}\ . 
\end{equation}

Returning to the Lagrangian description, we are looking for the state of lowest energy for given $O(2)$ charge and momentum.  Since the excitations lie entirely within an $O(2)$ this reduces to a free-field calculation.  Inserting the classical configuration
\begin{equation}
\vartheta = -\omega \tau + k \sigma\ , \quad \varphi_1 + i \varphi_2 = e^{i \vartheta}\ ,
\label{eq:vartheta}
\end{equation}
we have
\begin{equation}
{\cal J} = \frac{\omega}{g^2}\ ,\quad {\cal P} = \frac{\omega k}{g^2 }\ ,\quad
{\cal E} = \frac{\omega^2 + k^2}{2 g^2}\ . \label{eq:scen}
\end{equation}
Comparing with the Bethe ansatz results gives
\begin{equation}
g^2 = \frac{\pi}{2 h(\chi)^2}\ .
\end{equation}
Thus this state fixes the dictionary between the parameter $g^2$ of the Lagrangian description and the parameter $\chi$ of the Bethe ansatz description.

The zero mode integral equation cannot be solved in closed form; it can be solved numerically or by expanding around large or small $\chi$.  The details of the large $\chi$ expansion are set aside to the next subsection.  The result is
\begin{equation}
\rho_{0}(\phi) = \frac{ {\cal J}_R}{2\pi\sqrt{\chi}}\sqrt{\frac{\chi + \phi}{\chi - \phi}} + \frac{ {\cal J}_L}{2\pi\sqrt{\chi}}\sqrt{\frac{\chi - \phi}{\chi + \phi}}\ , \label{eq:r0u1}
\end{equation}
which integrates to
\begin{equation}
{\cal J}_{0} = \frac{\sqrt{\chi}}{2}( {\cal J}_R +  {\cal J}_L)\ .
\end{equation}
Then $h(\chi) = {\sqrt{\chi}}/{2}$ to leading order, and
\begin{equation}
g^2 = \frac{2\pi}{\chi} \ ,
\end{equation}
in agreement with the one-loop result~(\ref{eq:onel}).
In particular, the large-$\chi$ limit is the weak-coupling (classical) limit of the sigma model.
Note that in this regime, the zero-modes carry nearly all the charge, while the right- and left-movers carry all the energy and momentum. 

It is clear from the form of the zero mode equation that as we reduce $\chi$ we reduce ${\cal J}_0$ and so $g^2$ increases monotonically.  In the string world-sheet theory the limit $g^2 \to 
\infty$ is particularly interesting, because it is dual to the free gauge theory.  However, in our model we do not reach this limit even as $\chi \to 0$.  In this limit the zero mode range goes to zero and ${\cal J}_0 \to 0$, giving $h(0) = 1$ and $g^2 = \pi/2$.  The role of this special value is not clear.  It corresponds to the Kosterlitz-Thouless point, where the vortex interaction becomes marginal.\footnote{We thank H. Saleur for pointing this out, and that it is also the point where an exact lattice solution exists~\cite{RS}.  In our world-sheet approach, it is relatively easy to expand around this coupling.}  However, it is not clear that this continuum model should have a Kosterlitz-Thouless transition.  The small-$\chi$ expansion is simple to carry out to many terms~\cite{MP}, and it seems to be convergent and to allow continuation to negative $\chi$.   However, it does not seem that we can reach the very interesting point $g = \infty$, even at $\chi \to -\infty$.  This is a puzzling artifact of this supergroup coset model, which probably has no relevance to the string theory.

Excited states, with gaps in the sequence of Bethe integers $n_j$, correspond to excitations of the free field $\vartheta$~(\ref{eq:vartheta}).  In the application to string theory these can be removed using the residual gauge freedom of the conformal gauge: we can always choose coordinates in which $\vartheta = - \omega \tau + k \sigma$.  Thus to describe the physical states of string theory we can restrict attention to the filled rapidity band.  The center of mass constraints $E = P = 0$ must still be satisfied.

\subsection{Large-$\chi$ expansion}

Here and in section~4 we will work out some of the details of the large-$\chi$ expansion of the Bethe ansatz equations.  For the zero mode equation~(\ref{eq:zmeq}), the source terms on the right are strongly peaked at the endpoints, and so we start by analyzing the behavior near one endpoint, say $+\chi$.  Defining $\phi = \psi + \chi$ and $R(\psi) = \rho_0(\phi)$, the equation goes in the limit to 
\begin{equation}
R(\psi) - \int_{-\infty}^{0}  K_0(\psi -\psi') R(\psi') \ d\psi'= g(\psi) \ , \quad \psi < 0\ ,
\end{equation}
where
\begin{equation}
K_0(\psi -\psi') = \frac{2}{4\pi^2 + (\psi - \psi')^2}\ ,\quad
g(\psi) = \frac{2 {\cal J}_R}{4\pi^2 + \psi^2} \ .
\end{equation}

Again, the solution is via the Wiener-Hopf method, as reviewed for example in the appendix to~\cite{JNW} and in~\cite{has2}.  
In Fourier space one can write
\begin{equation}
1 - \tilde K_0(\omega)  = \frac{1}{G_+(\omega) G_-(\omega)}\ ,
\end{equation}
where the functions $G_+(\omega)$ and $G_-(\omega)$ are holomorphic and novanishing in the upper and lower half-planes respectively, and approach 1 at large $\omega$ in these respective half-planes.  Here we have the particular form
\begin{equation}
\tilde g(\omega) = {\cal J}_R \tilde K_0(\omega) = {\cal J}_R ( 1 - G_+^{-1}(\omega) G_-^{-1}(\omega))\ . \label{eq:kg}
\end{equation}
The integral equation then takes the form
\begin{equation}
G^{-1}_+ G^{-1}_- \tilde R = {\cal J}_R ( 1 - G_+^{-1} G_-^{-1}) + X_+
\end{equation}
where $X_+$ is an unknown function that is holomorphic in the upper half-plane and approaches 0 asymptotically (this appears because the integral equation holds only for negative $\psi$).  Multiplying by $G_+$ and rearranging gives
\begin{equation}
G^{-1}_- \tilde R  + {\cal J}_R (G_-^{-1}-1) =  {\cal J}_R( G_+ - 1) + G_+ X_+\ .
\end{equation}
The left-hand side is holomorphic in the lower half-plane and approaches 0 asymptotically, and the right-hand side has the same property in the upper half-plane.  It follows that both sides vanish identically, and so
\begin{equation}
\tilde R (\omega) = {\cal J}_R (G_-(\omega) - 1) \ . \label{eq:appii}
\end{equation}
Explicitly for the kernel~(\ref{eq:kg})
\begin{equation}
G_-(\omega) = \frac{1}{\sqrt{2\pi i\omega}} \Gamma(1+i\omega)e^{i\omega-i\omega\ln(i\omega)}\ .
\end{equation}

To match onto the solution away from the endpoint we need only the small-$\omega$ behavior, giving
 \begin{equation}
R(\phi) = \frac{{\cal J}_R}{\pi\sqrt{2(\chi-\phi)}} + O((\chi-\phi)^{-3/2})\ . \label{eq:mat}
\end{equation}
In the bulk we thus look for a solution of the form
\begin{equation}
\rho_0(\phi) = \frac{1}{{\chi}} r(\phi/\chi)\ , \label{eq:rphi}
\end{equation}
where
\begin{equation}
r(y) \sim \frac{{\cal J}_R \sqrt{\chi}}{\pi\sqrt{2(1-y)}}\ ,\quad y \to 1\ ,
\end{equation}
and similarly
\begin{equation}
r(y) \sim \frac{{\cal J}_L \sqrt{\chi}}{\pi\sqrt{2(1+y)}}\ ,\quad y \to -1\ .
\end{equation}
The large-$\chi$ limit of the zero mode equation~(\ref{eq:zmeq}) at fixed $y = \psi/\chi$ is
\begin{equation}
\dashint_{-1}^1 \frac{r(y')}{(y-y')^2} dy' = 0\ .
\end{equation}
We will discuss such principal part equations at more length in section~4.  Here there is a unique solution with the given limits,
\begin{equation}
r(y) = \frac{{\cal J}_R \sqrt{\chi}}{2\pi} \sqrt{\frac{1+y}{1-y}}
+
\frac{{\cal J}_L \sqrt{\chi}}{2\pi} \sqrt{\frac{1-y}{1+y}}\ ,
\end{equation}
giving eq.~(\ref{eq:r0u1}).  In section~2.2 we integrated this to obtain the zero mode charge density.  The full form~(\ref{eq:appii}) near the endpoint gives a correction to ${\cal J}_0$ that is subleading at large $\chi$.

This method can be iterated to give higher orders in the semiclassical expansion, but this is beyond our present scope.

\sect{The $SU(2)$ sector}

\subsection{The nested Bethe ansatz}

We now consider states with particles created by either $\varphi_1 + i \varphi_2$ or $\varphi_3 + i \varphi_4$.  That is, the particles of positively charged under
one of the factors in  
$O(2)\times O(2) \subset O(2m+2) \subset OSp(2m+2|2n)$.  In such states, the $E$ tensor still vanishes, but the $I$ and $P$ tensors are distinguishable.  The effective S-matrix at general $n$ is
\begin{equation}
S = \frac{i\theta P + \frac{2\pi}{n-2} I}{i\theta + \frac{2\pi}{n-2}} S_{pp}\ ,
\end{equation} 
where $S_{pp}$ is the single-species S-matrix~(\ref{eq:so2}).  A state with charge $J_1$ under the first $O(2)$ and $J_2$ under the second $O(2)$ is described in terms of $J=J_1 + J_2$ particles with $J_2$ impurities~\cite{yang}.
The resulting nested Bethe ansatz equations are
\begin{eqnarray} 
e^{ip_j L} &=& \prod_{\beta} \frac{i\theta_j - i\Lambda_{\beta} + \frac{\pi}{n - 2}}{i\theta_j - i\Lambda_{\beta} - \frac{\pi}{n - 2}} \prod_{i \ne j} S_{pp}(\theta_i - \theta_j)\ ,
\label{eq:rapbet}
\\
\prod_{j} \frac{i\theta_j - i\Lambda_{\alpha} + \frac{\pi}{n - 2}}{i\theta_j - i\Lambda_{\alpha} - \frac{\pi}{n - 2}} &=& \prod_{\beta \neq \alpha} \frac{i\Lambda_{\alpha} - i\Lambda_{\beta} - \frac{2\pi}{n - 2}}{i\Lambda_{\alpha} - i\Lambda_{\beta} + \frac{2\pi}{n - 2}}\ ,\label{eq:nested}
\end{eqnarray}
where indices $i,j$ run from $1$ to $J$ and $\alpha,\beta$ run from $1$ to $J_2$.  The $p_{j}$ and $\theta_{j}$ still describe the momenta and rapidity of each particle (including both types).
The pseudorapidities $\Lambda_\alpha$ describe the solution to a nested Bethe ansatz which describes the motion of the impurities on the chain of particles.  Eq.~(\ref{eq:nested}) gives a quantization condition for these $\Lambda_{\alpha}$, which are like spin chain rapidities.

\subsection{The single-impurity state}

Consider first a state
with a finite density of type 1 particles and a single type 2 particle.  We will use this in the next section to understand how the pseudorapidity $\Lambda$ maps onto the physical parameters.  
The Bethe ansatz equations reduce to
\begin{eqnarray}
e^{ip'_j L} &=& \frac{i\theta'_j - i\Lambda + \frac{\pi}{n - 2}}{i\theta'_j - i\Lambda - \frac{\pi}{n - 2}}\prod_{i \ne j}S(\theta'_i - \theta'_j) \ ,\label{eq:ba1}
\\
1 &=& \prod_{j} \frac{i\theta'_j - i\Lambda + \frac{\pi}{n - 2}}{i\theta'_j - i\Lambda - \frac{\pi}{n - 2}}\ . \label{eq:ba2}
\end{eqnarray}
Primes denote the rapidities and momenta in the single-impurity state, while the unprimed values refer to the no-impurity state, with $J$ particles all of type 1
as studied in the previous section.  

In the thermodynamic limit $L \rightarrow \infty$ the 
single-impurity state is treated as a perturbation of the pure state, defining $w(\theta) = L(\theta - \theta')$~\cite{thacker}.  Taking the logarithm of the Bethe ansatz equation~(\ref{eq:ba1}) and subtracting the unprimed equation gives
\begin{equation}
F(\theta) - \int_{-B_L}^{B_R} K(\theta - \theta')F(\theta') \, d\theta' = \frac{1}{\pi}\cot^{-1}\frac{(n - 2)(\theta - \Lambda)}{\pi}\ ,
\end{equation}
where $F(\theta) = w(\theta)\rho(\theta)$.  We have set $n_j = n'_j$, but leave the branch of the logarithm unspecified for now; different choices can be absorbed in shifts of the $n_j$. 

As we take the $n \rightarrow 2$ limit, the integral equation again splits into three parts.  The zero-mode equation is
\begin{equation}
\frac{1}{2} F_0(\phi) - \int_{-\chi}^{\chi} \frac{  F_0(\phi')}{4\pi^2 + (\psi - \psi')^2} d\phi' = \frac{1}{\pi}\cot^{-1}\frac{\phi - \tilde{\Lambda} }{\pi}\ , \label{eq:zper}
\end{equation}
where $\tilde{\Lambda} = (n - 2)\Lambda$; for states whose energy remains finite in the $n \to 2$ limit, it is $\tilde{\Lambda}$ that is held fixed.  The right- and left-moving parts can be put in the form 
\begin{eqnarray}
F_{R}(\theta) - \int_{-\infty}^{\tilde{B}_R} K_{RR}(\theta - \theta')F_R(\theta') \, d\theta' &=& \frac{1}{2}F_{0}(\chi)\ ,
\nonumber\\
F_{L}(\theta) - \int_{-\tilde{B}_L}^{\infty} K_{LL}(\theta - \theta')F_L(\theta') \, d\theta' &=& \frac{1}{2}F_{0}(-\chi) \label{eq:rlper}
\end{eqnarray}
We have used the zero mode equation to simplify these; note that $\cot^{-1}$ is essentially constant in the right and left ranges, and equal to its value at the nearer end of the zero-mode range.
Eqs.~(\ref{eq:rlper}) are solved readily using the Wiener-Hopf method to give
\begin{eqnarray}
\Delta P &=& \pi {\cal J}_R F_0(\chi) + \pi {\cal J}_L F_0(-\chi)\ , \label{eq:pper}
\\
\Delta E &=& \pi {\cal J}_R F_0(\chi) - \pi {\cal J}_L F_0(-\chi)\ . \label{eq:eper}
\end{eqnarray}
These are both functions of the rapidity $\tilde\Lambda$.  Eliminating $\tilde\Lambda$ gives the dispersion relation for $\Delta E$ in terms of $\Delta P$.

The second Bethe ansatz equation~(\ref{eq:ba2}) becomes
\begin{equation}
\frac{\hat m}{L} = \frac{1}{\pi}\int_{-\chi}^\chi F_0(\phi)\,\hat{\cot}^{-1}\frac{\phi - \tilde\Lambda}{\pi} \, d\phi\ . \label{eq:hatm}
\end{equation}
Here we have defined the $\hat{\cot}$ to vanish at $\pm \infty$ and to jump by $-\pi$ at 0, in parallel with eq.~(\ref{eq:hatln}).
Eq.~(\ref{eq:hatm}) provides a quantization condition on $\tilde\Lambda$ and so on $\Delta P$.  In fact, it follows immediately from taking the product of eq.~(\ref{eq:ba1}) over $j$ (so that $S(\theta_i'-\theta_j')$ cancels) that eq.~(\ref{eq:ba2}) for a single impurity directly implies quantization of momentum, $\Delta P = {2\pi m}/{L}$.  

\subsection{Equations for finite impurity density}

We now consider states with a finite density of both type 1 and type 2 particles.  
We assume that the rapidities $\theta_j$ lie in a single filled band and the pseudorapidities $\Lambda_\alpha$ lie in one or more filled bands.  The Bethe ansatz equations~(\ref{eq:rapbet},\,\ref{eq:nested}) then become the integral equations
\begin{equation}
\rho(\theta) -  \int_{-B_L}^{B_R} K(\theta - \theta')\rho(\theta') \, d\theta'
= \frac{M}{2\pi}\cosh\theta  - \int \frac{(n-2) \sigma(\Lambda) \, d\Lambda}{\pi^2 + (\theta - \Lambda)^2(n - 2)^2}
\end{equation}
and
\begin{equation}
\sigma(\Lambda) - 2\int \frac{(n - 2)\sigma(\Lambda') \ d\Lambda'}{4\pi^2 + (\Lambda - \Lambda')^2(n - 2)^2} = - \int_{-B_L}^{B_R} \frac{(n - 2)\rho(\theta) \, d\theta}{\pi^2 + (\theta - \Lambda)^2(n - 2)^2}\ .
\end{equation}
The pseudorapidity integral runs over the filled bands, which are not specified.  Each equation holds only within the filled range.
The total particle density is
\begin{equation}
{\cal J} = {\cal J}_1 + {\cal J}_2 = \int \rho(\theta) \, d\theta\ ,
\end{equation}
and the impurity density is
\begin{equation}
 {\cal J}_2 = \int \sigma(\Lambda) \, d\Lambda\ . \label{eq:impden}
 \end{equation}
 
 The $n \to 2$ limit is smooth if we define $\phi = (n-2)\theta$ and $\tilde\Lambda = (n-2) \Lambda$ as before, and $\tilde\sigma(\tilde\Lambda) = \sigma(\Lambda)/(n-2)$.  Then
 \begin{equation}
\frac{1}{2}\rho_{0}(\phi) -  \int_{-\chi}^{\chi} \frac{\rho_{0}(\phi') \ d\phi'}{4\pi^2 + (\phi - \phi')^2}
= \frac{{\cal J}_R}{4\pi^2 + (\chi - \phi)^2} + \frac{{\cal J}_L}{4\pi^2 + (\chi + \phi)^2}  - \int \frac{\tilde\sigma(\tilde\Lambda)\, d\tilde\Lambda}{\pi^2 + (\phi - \tilde\Lambda)^2}
\label{eq:rho0bet}
\end{equation}
and
\begin{equation} \label{eq: anyrange}
 \tilde\sigma(\tilde\Lambda)  - 2\int \frac{\tilde\sigma(\tilde\Lambda') \, d\tilde\Lambda'}{4\pi^2 + (\tilde\Lambda - \tilde\Lambda')^2} = 
-\frac{{\cal J}_R}{\pi^2 + (\chi - \tilde\Lambda)^2}
- \frac{{\cal J}_L}{\pi^2 + (\chi + \tilde\Lambda)^2}
- \int_{-\chi}^\chi 
\frac{\rho_0(\phi)}{\pi^2 + (\phi - \tilde\Lambda)^2}\ .
\end{equation}
The right- and left-moving equations (\ref{eq:rightleft}) are unchanged --- all additional terms scale out as $n\to 2$.  Thus the relation~(\ref{eq: PandE}) continues to hold, determining the energy and momentum in terms of ${\cal J}_{R,L}$.

The equations for $\rho_0$ and $\tilde\sigma$ are coupled, but they decouple if all impurities are at large pseudorapidity $|\tilde\Lambda| \gg \chi$.  We will see in the next section that this corresponds to nonrelativistic impurities.  The Bethe ansatz equations become
\begin{equation}
\frac{1}{2}\rho_{0}(\phi) -  \int_{-\chi}^{\chi} \frac{\rho_{0}(\phi') \ d\phi'}{4\pi^2 + (\phi - \phi')^2}
= \frac{{\cal J}_R}{4\pi^2 + (\chi - \phi)^2} + \frac{{\cal J}_L}{4\pi^2 + (\chi + \phi)^2}  - \frac{\cal T}{\chi^2} \label{eq:rhodec}
\end{equation}
and
\begin{equation} \label{eq: anyrangeNR}
\tilde\sigma(\tilde\Lambda) - 2\int \frac{\tilde\sigma(\tilde\Lambda') \, d\tilde\Lambda'}{4\pi^2 + (\tilde\Lambda - \tilde\Lambda')^2}  = 
-\frac{{\cal J}}{\pi^2 + \tilde\Lambda^2}\ ,
\end{equation}
where
\begin{equation}
{\cal T} = \chi^2 \int \frac{\tilde\sigma(\tilde\Lambda) \,d\tilde\Lambda}{\pi^2 + \tilde\Lambda^2}\ .
\end{equation}

\sect{The large-$\chi$ approximation}

The integral equations found in section~3 determine the world-sheet energies in the quantized world-sheet theory.  In order to make contact with earlier results, we now take the large-$\chi$ approximation, which we have seen to be the classical limit of the field theory.  For reference recall our semiclassical result for $\chi(g^2)$, and express it in terms of the string theory quantities:
\begin{equation}
\chi = \frac{2\pi}{g^2} = \frac{R_{\rm AdS}^2}{\alpha'} =  \lambda^{1/2}\ .
\end{equation}

\subsection{The single impurity}

For $\ell = \tilde{\Lambda}/\chi > 1$, the argument of the $\cot^{-1}$ becomes large and negative in the semiclassical limit.  It is convenient to choose the branch $-\pi < \cot^{-1} < 0$ so that the inhomogenous term in the zero-mode equation is small, 
\begin{equation}
\frac{1}{2}F_0(\phi) - \int_{-\chi}^{\chi} \frac{F_0(\phi') \ d\phi'}{4\pi^2 + (\phi - \phi')^2} = \frac{1}{\phi - \tilde{\Lambda}} \ .
\end{equation}
As $\chi \to \infty$, there are two ways to take the limit: we can make a linear shift of $\phi$ to focus on the behavior one or the other endpoint, or we can make a multiplicative rescaling of $\phi$ to keep the range finite.  In practice it is necessary to do both and match the solutions.  In 
section~2.3 the source was peaked at the endpoints and so we analyzed the endpoint behavior
first.  Here it is distributed and we analyze the bulk behavior first.  

Defining 
\begin{equation}
y = \phi/\chi\ ,\quad \ell = \tilde\Lambda/\chi\ ,\quad f(y) = F_0(\phi) \ ,
\end{equation}
the zero mode equation has the large-$\chi$ limit
\begin{equation}
- \dashint_{-1}^{1} \frac{f(y) \ dy '}{ (y - y')^2} = \frac{1}{y - \ell} \ .
\end{equation}
The principal part arises because the $\frac{1}{2}F_0$ just cancels the area under the peak in the integral.  The solution is given in eq.~(\ref{eq:iiisol}).  In particular the limits~(\ref{eq:iiilims}) are
\begin{eqnarray}
F_0(\phi \sim \chi) &=& -\frac{\sqrt 2}{\pi} \frac{\sqrt{\chi-\phi}}{\sqrt\chi} \Biggl( \frac{\sqrt{\ell+1}}{\sqrt{\ell-1}} - 1 \Biggr)\ ,
\nonumber\\
F_0(\phi \sim \chi) &=& -\frac{\sqrt 2}{\pi} \frac{\sqrt{\chi+\phi}}{\sqrt\chi} \Biggl( 1 - \frac{\sqrt{\ell-1}}{\sqrt{\ell+1}} 
\Biggr)\ . \label{eq:F0lims}
\end{eqnarray}

The principle part approximation to the Bethe equation breaks down when the distance from $\phi$ to an endpoint $\pm\chi$ is of order one.  The momentum and energy shifts~(\ref{eq:pper},\,\ref{eq:eper}) depend on the value {\it at} the endpoint, and so we need to work out the endpoint correction.  For $\phi -\chi$ of order one define
\begin{equation}
\psi = \phi - \chi\ , \quad \Phi(\psi) = \sqrt{\chi} F_0(\phi)\ .
\end{equation}
Matching to the bulk solution, we see that at large negative $\psi$ $\Phi(\psi) \to c |\psi|^{1/2}$ with a known coefficient.  Inserting this form into the Bethe equation gives
\begin{equation}
\Phi(\psi) - \int_{-\infty}^{0} \frac{2 \Phi(\psi') \, d\psi'}{4\pi^2 + (\psi - \psi')^2} = O(\chi^{-1/2})\to 0\ .
\label{eq:Phi}
\end{equation}
Thus we need to solve the sourceless equation with give large-$\psi$ behavior.  We cannot immediately apply the Wiener-Hopf method because the Fourier transform does not exist.  Thus we differentiate eq.~(\ref{eq:Phi}) once to obtain
\begin{equation}
\Upsilon(\psi) - \int_{-\infty}^{0} \frac{2\Upsilon(\psi') \, d\psi'}{4\pi^2 + (\psi - \psi')^2} 
= - \frac{2 \Phi(0)}{4\pi^2 + \psi^2 }
\end{equation}
for $\Upsilon(\psi) = \Phi'(\psi)$.  This is now of the same form as encountered in section~2.3.  In particular, to match onto the bulk equation we need the asymptotic form~(\ref{eq:mat}),
\begin{equation}
\Upsilon(\psi) \to -\frac{\Phi(0)}{\pi\sqrt{2 \psi}} + O(\psi^{-3/2})\ ,
\quad
\Phi(\psi) \to -\frac{\sqrt{2 \psi}}{\pi} \Phi(0) + O(\psi^{-1/2})\ .
\end{equation}
Matching onto eq.~(\ref{eq:F0lims}) gives the necessary result
\begin{equation}
F(\chi) = \frac{1}{\sqrt{\chi}} \Biggl( \frac{\sqrt{\ell+1}}{\sqrt{\ell-1}} - 1 \Biggr)
\ ,\quad
F(-\chi) = \frac{1}{\sqrt{\chi}} \Biggl(1-\frac{\sqrt{\ell-1}}{\sqrt{\ell+1}} \Biggr)
\ .
\end{equation}

Taking for simplicity the case ${\cal J}_R = {\cal J}_L =  {\cal J}/\sqrt{\chi}$ (the second equality is the already-known large-$\chi$ result), the impurity energy and momentum become
\begin{equation}
\Delta P = \mu \frac{1}{\sqrt{\ell^2-1}}\ , \quad
\Delta E = \mu \Biggl( \frac{\ell}{\sqrt{\ell^2-1}} - 1 \Biggr)
\ ,\quad
\mu = 2\pi {\cal J}/\chi\ .
\end{equation}
This can be put in the form of a relativistic dispersion relation 
\begin{equation}
\left(\Delta E + \mu \right)^2 - \Delta P^2 = \mu^2 \ ,\quad
\Delta P > 0\ . \label{eq:reldis}
\end{equation}
The momentum quantization condition gives
\begin{equation}
\ell = \sqrt{ \frac{\mu^2 L^2}{(2\pi \hat m)^2} - 1 }\ . \label{eq:hatm2}
\end{equation}
For $\ell < -1$ it is simplest to take the branch $0 < \cot^{-1} < \pi$, giving
\begin{equation}
\Delta P = - \mu \frac{1}{\sqrt{\ell^2-1}}\ , \quad
\Delta E = \mu \Biggl( \frac{|\ell|}{\sqrt{\ell^2-1}} - 1 \Biggr)\ , \end{equation}
which corresponds to the $\Delta P < 0$ branch of the relativistic dispersion relation~(\ref{eq:reldis}).  

For $-1 < \ell < 1$ we approximate
\begin{equation}
\frac{\chi}{\pi}\tan^{-1}\frac{\pi}{\chi(y - \ell)} = -\chi(p + \Theta(y - \ell))
\end{equation}
where $\Theta(x)$ is the step function and $p$ is an integer associated with the branch choice for $\tan^{-1}$.  Here, there is no obvious preference between $p = 0$ and $p = -1$, so we will leave it undetermined.  Then
\begin{eqnarray}
F_{0}(\chi) &=& \frac{\chi}{\pi}\left[p\pi + \frac{\pi}{2} + \sqrt{1 - \ell^2} - \tan^{-1}\frac{\ell}{\sqrt{1 - \ell^2}}\right]\ ,
\nonumber\\
F_{0}(-\chi) &=& \frac{\chi}{\pi}\left[p\pi + \frac{\pi}{2} - \sqrt{1 - \ell^2} - \tan^{-1}\frac{\ell}{\sqrt{1 - \ell}}\right]\ ,
\end{eqnarray}
and
\begin{eqnarray}
\Delta P &=& {\cal J}\Biggl((2p + 1)\pi - 2\tan^{-1}\frac{\ell}{\sqrt{1 - \ell^2}}\Biggr)\ ,
\nonumber\\
\Delta E &=& 2{\cal J} \sqrt{1 - \ell^2}\ .
\end{eqnarray}
and this gives
\begin{equation}
\Delta E = 2{\cal J}\left|\sin ({\Delta P}/{2{\cal J}})\right|, \ \ \ \ \ 2\pi np < \Delta P < 2\pi n(p + 1)\ .
\end{equation}

Let us compare with the semiclassical calculation in the field theory.
We focus on an $O(4) \subset O(2m+2)$ subgroup,
\begin{equation} \label{eq: action}
S = -\frac{1}{2g^2}\int d\tau\, d\sigma\, \partial_{\mu} X_{i}\partial^{\mu}X_{i}\ ,
\end{equation}
with $i = 1,2,3,4$ and the constraint $X_{i}X_{i} = 1$.  We transform to variables $X_{3,4}$ and $\phi$, where
\begin{equation}
X_{1} = \cos\phi \sqrt{1 - X_{3}^2 - X_{4}^2}\ ,\quad
X_{2} = \sin\phi \sqrt{1 - X_{3}^2 - X_{4}^2}\\ .
\end{equation}
The perturbation transforms as $X_3 + i X_4$, so we expand to quadratic order in $X_{3,4}$ to obtain the Hamiltonian
\begin{equation}
{H} = \int d\sigma\, \left\{ \frac{g^2}{2}(\pi_{3}^2 + \pi_{4}^2 + \pi_{\phi}^2) + \frac{g^2}{2}\pi_\phi^2 (X_{3}^2 + X_{4}^2) + \frac{1}{2g^2}( X_{3}'^2 + X_{4}'^2 + \phi'^2) \right\} \ .
\end{equation}
Note that $\pi_\phi = \cal J$ in the unperturbed state, so $X_{3,4}$ indeed behave as relativistic particles of mass $\mu = g^2 {\cal J} = 2\pi {\cal J}/\chi$.  Removing one $X_{1+i2}$ charge from the sea and adding one $X_{3+i4}$ particle of momentum $\Delta P$ thus changes the energy by
\begin{equation}
\Delta E = \sqrt{\Delta P^2 + \mu^2} - \mu
\end{equation}
as found above; the $-\mu$ term is from $\pi_\phi L \to \pi_\phi L - 1$.

The semiclassical calculation covers the ranges $|\lambda| > 1$ only.  As $|\lambda| \to 1$ the energy becomes large and apparently the semiclassical description breaks down.  The need to take different branches of the $\cot^{-1}$ for $\ell > 1$ and $\ell < -1$ reflects an interesting spectral flow phenomenon.  If we start with large positive $\ell$ and move to decreasing values, we have increasing positive momentum.  If we decrease $\ell$ through zero and then past $-1$ while remaining on the original branch of the $\cot^{-1}$, we reach a state with an impurity of {\it negative} momentum.  However, the total momentum of the state must increase throughout, because $\Delta P = 2\pi \hat m/L$ is increasing monotonically with $m$.  The point is that the $\cot^{-1}$ approaches a constant value $-\pi$, which reflects a shift of the momenta of the sea particles, an increase of one unit of momentum for each.  
The results for $-1 < \lambda < 1$ suggest a simple interpretation: as the impurity pseudorapidity passes through the sea a hole appears, with all particles at $y > \ell$ shifted one unit to the right.  When the impurity reaches $\ell=-1$ the whole sea is shifted, giving total momentum $2\pi J/L$.  The energy shift at this point is of higher order in $1/L$.

\subsection{Nonrelativistic impurities}

Now consider a finite density of nonrelativistic impurities.  From the single-impurity example we see that these are at $|\ell| \gg 1$.  We thus have the Bethe equations given at the end of section~3.3.  

The $\rho_0$ equation differs from the earlier~(\ref{eq:zmeq}) by the constant term $- {\cal T} /\chi^2$.
Defining
\begin{equation}
y= \phi/\chi\ ,\quad r(y) = \rho_0(\phi) {\chi}
\end{equation}
as in section~2.3, the Bethe equation in the bulk becomes
\begin{equation}
\dashint_{-1}^1 \frac{r(y')}{(y-y')^2} dy' = {\cal T} \ . \label{eq:betr}
\end{equation}
Eq.~(\ref{eq:semi}) then gives the additional contribution
\begin{equation}
\rho_{0}(\phi) = \frac{ {\cal J}_R}{2\pi\sqrt{\chi}}\sqrt{\frac{\chi + \phi}{\chi - \phi}} + \frac{ {\cal J}_L}{2\pi\sqrt{\chi}}\sqrt{\frac{\chi - \phi}{\chi + \phi}} -\frac{\cal T}{\chi^2\pi}\sqrt{\chi^2 - \phi^2}\ .
\end{equation}
This integrates to
\begin{equation}
{\cal J}_{0} = \frac{\sqrt{\chi}}{2}( {\cal J}_R +  {\cal J}_L) - \frac{ {\cal T} }{2}\ .
\end{equation}
The shift of ${\cal J}_0$ and of the energy if of order $\ell^{-2}$.  Expanding to second order in $\ell^{-1}$ and ${\cal P}$, the energy~(\ref{eq:energ}) becomes
\begin{equation}
\mathcal{E} =\frac{\pi{\cal J} ^2}{\chi} 
+  \frac{ \mathcal{P}^2\chi}{4\pi{\cal J} ^2} +
\frac {\pi {\cal T}{\cal J} }{ \chi} \ . \label{eq:nre}
\end{equation}

The $\tilde\sigma$ equation has a smooth source and so we take the bulk limit,
\begin{equation}
\ell = \tilde\Lambda/\chi\ ,\quad s(\ell) = \chi \tilde\sigma(\tilde\Lambda)\ .
\end{equation}
The Bethe equations become
\begin{equation}
\dashint \frac{s(\ell')\, d\ell'}{(\ell - \ell')^2} = \frac{\cal J}{2\ell^2}\ ,
\label{eq:sbet}
\end{equation}
where again the contours are unspecified, and might even be continued into the complex $\ell$ plane.
This density feeds back into the energy~(\ref{eq:nre}) through
\begin{equation}
{\cal T} =  \int \frac{s(\ell)\, d\ell}{\ell^2}\ .
\end{equation}
In this case we do not need a separate analysis of the endpoint region, because its effect on $\cal T$ is subleading in $\chi$.  The last term in the energy density is then
\begin{equation}
\Delta {\cal E} = \frac{\mu}{2} \int \frac{s(\ell)\, d\ell}{\ell^2}\ .
\end{equation}
It is useful to integrate eq.~(\ref{eq:sbet}) to obtain
\begin{equation}
\dashint \frac{s(\ell')\, d\ell'}{\ell - \ell'} = \frac{\cal J}{2\ell}
-\frac{\chi\hat m}{2 L}\ , \label{eq:intsbet}
\end{equation}
where $\hat m$ must be constant on each connected band of impurities, and in fact must be an integer by the Bethe ansatz equation~(\ref{eq:nested}).  For a small number of impurities it is the same as $\hat m$ in the single impurity eqs.~(\ref{eq:hatm},\,\ref{eq:hatm2}).

To impose the analog of the physical state equations from string theory, we need also the integrated form of eq.~(\ref{eq:betr}),
\begin{equation}
\dashint_{-1}^1 \frac{r(y')}{y-y'} dy' = - \int \frac{s(\ell)\, d\ell}{\ell}
+ \frac{\chi \hat l}{L} \label{eq:hatl}
\end{equation}
where $\hat l$ is an integer.  (To derive this one must integrate the Bethe ansatz equation before taking the classical limit).
The physical state condition $P = 0$ implies, by the general result~(\ref{eq: PandE}), that ${\cal J}_R = {\cal J}_L$.  The left-hand side of eq.~(\ref{eq:hatl}) then vanishes, and so we have the constraint
\begin{equation}
\int \frac{s(\ell)\, d\ell}{\ell} = \frac{\chi \hat l}{L}\ . \label{eq:scon}
\end{equation}
Finally, to satisfy the physical state condition $E_{\rm total} = 0$ we append a free timelike field,
whose energy is like the classical result~(\ref{eq:scen}) but with a minus sign,
\begin{equation}
{\cal E} = - \frac{g^2 }{2}{\cal D}^2\ ,\quad {\cal D} = \Delta/L\ , 
\end{equation}
where $\Delta$ is the spacetime dimension.  In all,
\begin{equation}
0 = \frac{g^2 }{2}({\cal J}^2 - {\cal D}^2) + \frac{g^2 {\cal J}}{2} \int \frac{s(\ell)\, d\ell}{\ell^2}\ , \label{eq:oneldim}
\end{equation}
where we have used $g^2 = 2\pi/\chi$.

The Bethe equations~(\ref{eq:intsbet},\,\ref{eq:scon},\,\ref{eq:oneldim}) are the same as 
in the nonrelativistic classical limit of the sigma model, which reproduces the one-loop anomalous dimensions of the gauge theory; see ref.~\cite{KMMZ} for a detailed discussion.  Note that the nonrelativistic expansion parameter is $k^2/\mu^2$ where $k = 2\pi \hat m/L$ is the wavenumber of the impurity on the string.  The expansion parameter reduces to $(2\pi \hat m)^2
/g^4 J^2 = \hat m^2 \lambda/J^2$, and so for fixed harmonic $\hat m$ the nonrelativistic expansion it is the same as the dual gauge theory loop expansion~\cite{FT2}.  The agreement is expected, because the leading large-$\chi$ approximation reduces to the $SU(2)$ sector of the bosonic sigma model, which is the same here as in the string theory.  It confirms that the $n \to 2$ limit that we are considering gives a sensible Bethe ansatz, and shows one way that these can be extended to a quantized sigma model, eqs.~(\ref{eq:rhodec},\,\ref{eq: anyrangeNR}).

While on the subject of the nonrelativistic limit, we should note that even the quantum-mechanical equations can be simplified in this limit.  The point is that the pseudorapidities $\tilde\Lambda$ are much larger than 1, so for finite impurity density the impurity bands have length much greater than 1.  The $\tilde\sigma$ equation can then be reduced to the same principal part equation~(\ref{eq:intsbet}) and therefore the moment ${\cal T}$ is unchanged.  However, the contour for the $\rho_0$ equation is not long, so the relation between ${\cal T}$ and the energy (and dimension) will be corrected.  It follows that states that have equal dimensions in the nonrelativistic semiclassical limit still have equal dimensions in the nonrelativistic quantum theory.  This is similar to the string world-sheet theory result but somewhat weaker, for in that case no $g^2$ correction is expected at all~\cite{FT}.  In both cases the system should be described by a low-energy effective action for the impurities~\cite{FT}.  In the string theory case this is not renormalized, whereas in our less supersymmetric model there is evidently a renormalization of the parameters.  Note however that even in more supersymmetric theories one expects nonrenormalization results to become weaker as one goes to higher dimension operators, which may be connected with the three-loop discrepancy of refs.~\cite{Cetal,SS}.

\subsection{The general case}

We now consider the large-$\chi$ limit without assuming nonrelativistic impurities.  The equations 
for $\rho_0$ and $\tilde\sigma$ are now coupled, but we expect by analogy with ref.~\cite{KMMZ} to be able to obtain an equation for $\tilde\sigma$ by itself.  We will do this by solving for $\rho_0$.  

The Bethe equation~(\ref{eq:rho0bet}) in linear in $\rho_0$ and has three source terms on the right.  We separate
\begin{equation}
\rho_0(\phi) = \OL{\rho}_0(\phi) + \Delta r(y)/\chi\ , \quad y = \phi/\chi\ ,
\end{equation}
where $\OL{\rho}_0$ is sourced by the first two terms and $\Delta r$ by the third.  Then $\OL{\rho}_0$ is exactly the same as for the $U(1)$ sector,
\begin{equation}
\OL{\rho}_0 = \frac{ \tilde{\cal J} \sqrt{\chi}}{\pi\sqrt{\chi^2 - \phi^2}}\ ,
\end{equation}
where we have used the physical state condition ${\cal J}_R = {\cal J}_L \equiv \tilde {\cal J}$.  For $\Delta r$ and $s$ we obtain the principal part equations
\begin{eqnarray}
\dashint_{-1}^1 \frac{\Delta r(y') \,  dy'}{(y-y')^2} &=&  \int \frac{s(\ell)\, d\ell}{(y - \ell)^2}\ ,
\nonumber\\
2\dashint \frac{s(\ell') \ d\ell'}{(\ell - \ell')^2} &=& \frac{\tilde{\cal J}}{(\ell - 1)^2} + \frac{\tilde{\cal J}}{(\ell + 1)^2} + \int_{-1}^{1} \frac{\chi {\rho}_0(\chi y) \, dy}{(\ell - y)^2}\ .
\end{eqnarray}
The integrated forms are
\begin{eqnarray}
\dashint_{-1}^1 \frac{\Delta r(y') \,  dy'}{y-y'} &=&  \int \frac{s(\ell)\, d\ell}{y - \ell}
+ \frac{\chi \hat l}{L}\ ,
\label{eq:deltrbet}\\
2\dashint \frac{s(\ell') \ d\ell'}{\ell - \ell'} &=& \frac{\tilde{\cal J}}{(\ell - 1)} + \frac{\tilde{\cal J}}{(\ell + 1)} + \int_{-1}^{1} \frac{\chi {\rho}_0(\chi y)  \, dy}{\ell - y} -\frac{\chi\hat m}{L}
\nonumber\\
&=& \frac{ \tilde {\cal J} \sqrt{\chi}}{\sqrt{\ell^2 - 1}}
 + \int_{-1}^{1} \frac{\Delta r(y)  \, dy}{\ell - y} -\frac{\chi\hat m}{L}
\ , \label{eq:sbet2}
\end{eqnarray}
dropping a term of relative order $\chi^{-1/2}$.

We now solve eq.~(\ref{eq:deltrbet}) for $\Delta r$, using the inverse finite Hilbert transform~(\ref{eq:simpgen}).  Note that the solution exists only with the constraint~(\ref{eq:fhcon}),
which becomes
\begin{equation}
\int \frac{s(\ell)\, d\ell}{\sqrt{\ell^2 - 1}} = \frac{\chi \hat l}{L}\ . \label{eq:scon2}
\end{equation}
The solution is then
\begin{equation}
\Delta r(y) =  \frac{\sqrt{1 - y^2}}{\pi}\int \frac{s(\ell) \, d\ell}{(y - \ell)\sqrt{\ell^2 - 1}}\ .
\end{equation}
Substituting back into the eq.~(\ref{eq:sbet2}) gives
\begin{equation}
\dashint \frac{s(\ell') \, d\ell'}{\sqrt{\ell'^2 - 1}}\left[\frac{\sqrt{\ell^2 - 1} + \sqrt{\ell'^2 - 1}}{\ell - \ell'}\right] = \frac{\tilde{\cal J}\!\sqrt{\chi}}{\sqrt{\ell^2 - 1}}
+ \frac{\chi (\hat l - \hat m)}{L}\ ,
\end{equation}
again dropping a term of relative order $\chi^{-1/2}$.

As desired, we have found a closed equation for $s(\ell)$, but of a somewhat complicated form.   The equation simplifies if we make the change of variables
\begin{equation}
\ell = \frac{x^2 + 1}{2x}\ , \quad \eta(x) = \frac{2\pi L}{\chi} s(\ell)\ .
\end{equation}
This is the same change of variables used to relate the rapidities in the gauge description to the spectral parameter of the monodromy matrix~\cite{KMMZ,BDipS};
note that $\eta(x)$ is not a density~\cite{BDipS}.  Note also that the $x$ plane is mapped to two copies of the $\ell$ plane, through a cut between $-1$ and $1$.
The Bethe equation becomes
\begin{equation}
2\dashint \frac{\eta(x') \, dx'}{x - x'} = \frac{2\pi L \tilde{\cal J}}{(x - 1) \sqrt{\chi}} + \frac{2\pi L  \tilde{\cal J}}{(x + 1) \sqrt{\chi}} - {2\pi \hat m}\ .
\end{equation} 
This is the general classical Bethe equation found in ref.~\cite{KMMZ}.  As was shown in that work, the various folded and spinning solutions can be obtained from it.

To complete the comparison we relate the various constants to moments of $\eta$.  The constraint~(\ref{eq:scon2}) becomes
\begin{equation}
\int \frac{\eta(x)\, dx}{x} = 2\pi \hat l \ . \label{eq:scon3}
\end{equation}
The total particle density is
\begin{equation}
{\cal J} = 2\tilde{\cal J} + \int_{-\chi}^\chi \rho_0(\phi)\,  d\phi = \tilde{\cal J} \!\sqrt{\chi}
- \frac{\chi}{2\pi L}\int \frac{\eta(x) \, dx}{x^2} \ ,
\end{equation}
dropping a term of relative order $\chi^{-1/2}$.  Defining the dimension as insection~4.2, we have the general result $0 = E_{\rm total} / L= \pi \tilde {\cal J}^2 - g^2 {\cal D}^2/ 2$, and so at large $\chi$
\begin{equation}
{\cal D} = \tilde{\cal J} \! \sqrt{\chi} = {\cal J} + \frac{\chi}{2 \pi L}\int \frac{\eta(x) \, dx}{x^2}\ .
\end{equation}
The number density of type 2 particles is given by the integral over the pseudorapidity density, 
\begin{equation}
{\cal J}_{2} = \frac{\chi}{4\pi L}\int \eta(x)\Biggl[1 - \frac{1}{x^2}\Biggr] \, dx\ .
\end{equation}

These results are equivalent to eqs. 4.43, 4.44, 4.45, and 4.47 of \cite{KMMZ}, with the notation
$\eta \to \rho$, $L \to 2\pi$, $L{\cal J} \to L$, $L {\cal J}_2 \to J$, $L{\cal D} \to \Delta$, $\hat l \to m$, $\hat m \to -n$.

\sect{Discussion}

Let us first review the expansion parameters for the various approximations.  For the nonrelativistic approximation it is
$\hat m^2  \lambda / J^2$.  For the finite size expansion it is $1/J$.  For the world-sheet quantum field theory it is $\lambda^{-1/2}$.  Thus the expansion for $ \Delta$, assuming that it is analytic in all the parameters, is
\begin{equation}
\Delta = { J} \sum_{a,b,c = 0}^\infty c_{abc} \Biggl( \frac{\hat m^2  \lambda }{ J^2 } \Biggr)^a \Biggl( \frac{1}{ J} \Biggr)^b 
\lambda^{-c/2}\ .
\end{equation}
If we consider only the $J$ and $\lambda$ dependence there are degeneracies.  Increasing $b$ by two is the same as increasing $a$ by one and $c$ by two.  However, these are distinct physical effects.  For example, they can be distinguished by their $\hat m$ dependence.  If we take $\hat m$ and $J$ to infinity together with the ratio fixed, it amounts to taking the length $L$ to infinity with fixed world-sheet wavelength.  In this limit the 
$\lambda^{-1/2}$ effects dominate the $1/J$ effects.
In our model we believe that our integral equations capture the full world-sheet quantum theory, but no finite size effects at present. 

We have developed techniques for deriving and solving the Bethe ansatz in conformal world-sheet theories.  The unexpected zero modes played an interesting role.  In the nonrelativistic limit we were able to decouple them from the impurities, though they themselves still had nontrivial quantum Bethe equations.  In the semiclassical limit we were able to solve and eliminate them.  This had the interesting effect of introducing a cut in the rapidity plane, which was removed by changing to the monodromy variable.  In the fully relativistic quantum theory we do not know how to solve for the zero modes analytically, and it may indeed be necessary to retain this additional degree of freedom.

To complete the solution of the planar ${\cal N}=4$ theory it is necessary to understand both the $\lambda^{-1/2}$ and the $1/J$ effects.  Perhaps the powerful and elegant approach of ref.~\cite{KMMZ} can be extended directly.  In our approach, inclusion of the $\lambda^{-1/2}$ effects would require that we find an S-matrix for the $AdS_5 \times S^5$ world-sheet theory, either by the limit from a massive integrable theory or directly.  The principal chiral supergroup models may be more similar in structure to the $AdS_5 \times S^5$ theory and give some insight.  For the finite-size effects, the most direct approach would be to identify a bare version of the theory with a `ferromagnetic' state.  Such models exist for some bosonic cosets~\cite{PW,Wieg,FadR}, but it is not clear whether the extension to a supergroup symmetry is possible.

In summary, there is still every reason to expect that a Bethe ansatz solution exists for the full planar ${\cal N} = 4$ theory, but there remain some important hurdles.  We believe that our work points to an important gap in the current understanding, namely the $\lambda^{-1/2}$ quantum effects, and gives some indication as to how these are to be included.

\section*{Acknowledgments}

We would like to thank N. Beisert, T. Erler, A. Ludwig, R. Roiban, H. Saleur, M. Staudacher, and A. Volovich for useful discussions.  This work was supported by National Science Foundation
grants PHY99-07949 and PHY00-98395. The work of N.M. was also supported by a
National Defense Science and Engineering Graduate Fellowship.

\appendix

\sect{Appendix: Principal part equations}

The large-$\chi$ limit of the Bethe equations lead to the integral equations of the form
\begin{equation}
\dashint_{a}^{b} \frac{f(y')}{(y-y')^2}\, dy = j(y) \label{eq:bulkg}
\end{equation}
or its integral with respect to $y$
\begin{equation}
\dashint_{a}^{b} \frac{f(y') \, dy'}{y - y'} = V(y)\ , \label{eq:simp}
\end{equation}
where $V'(y) = -j(y)$.  The additive constant in $V(y)$ is undetermined by this definition, but we will see that it is determined by the integral equation.  These are finite Hilbert transforms and their inversion is well-known.  We will work out both the general form and some useful special cases.

Eq.~(\ref{eq:simp}) arises in the evaluation of matrix integrals, e.g.~\cite{DfGZj}; we repeat here the method of solution for convenience.  Define for complex $z$ the function
\begin{equation}
g(z) = \int_{a}^{b} \frac{f(y') \ dy'}{z - y'}
\end{equation}
so that
\begin{equation}
V(y) = \frac{1}{2}\left[g(y + i\epsilon) + g(y - i\epsilon)\right]\ ,\quad
f(y) = \frac{1}{2\pi i}\left[g(y - i\epsilon) - g(y + i\epsilon)\right]\ .
\end{equation}
Then
\begin{eqnarray}
g^2(z) &=& \dashint_{a}^{b}\dashint_{a}^{b} \frac{f(y') f(y'') \ dy' \ dy''}{(z - y')(z - y'')} \nonumber \\
& = & \dashint_{a}^{b}\dashint_{a}^{b} f(y') f(y'') \ dy' \ dy'' \ \left[\frac{1}{z - y'} - \frac{1}{z - y''}\right]\frac{1}{y' - y''} \nonumber\\
& = & 2\dashint_{a}^{b} \frac{f(y') \ dy'}{z - y'} V(y') \nonumber\\
& = & 2\dashint_{a}^{b} \frac{f(y') \ dy'}{z - y'}\left[V(y') - V(z)\right] + 2V(z)g(z)\ . \label{eq:h2}
\end{eqnarray}
We will eventually use this to derive a general Green's function solution to eq.~(\ref{eq:bulkg}), but first obtain some simple special solutions.

\subsubsection*{Special case I: $j(y) = 1$.}  We have $V = -y + C$, and eq.~(\ref{eq:h2}) becomes
\begin{equation}
g^2(z) = \kappa + 2 V(z) g(z)\ ,\quad \kappa= 2  \int_a^b f(y)\,dy\ ,
\end{equation}
or
\begin{equation}
g(z) = - z + C  + \sqrt{(C -  z)^2 + \kappa}\ .
\end{equation}
The branch of the square root, here and below, is fixed by the property $g(z) \to 0$ as $z \to \infty$.  From its definition, $g(z)$ has a branch cut on the real line from $a$ to $b$, which determines $C =  (a+b)/2 $ and $\kappa = -(b-a)^2/4$.  Thus $g(z) = - z + C +  \sqrt{(z-a)(z-b)}$, and
\begin{equation}
f(y) = -\frac{1}{\pi}\sqrt{(y-a)(b-y)}\ . \label{eq:semi}
\end{equation}
This is the Wigner semi-circle law for eigenvalues of gaussian-random matrices.

\subsubsection*{Special case II: $j(y) = 1/(y - z_{0})^2$.}   

 The value $z_0$ may be complex but is assumed not to lie directly in the integration range $(a,b)$.  Here
\begin{equation}
V(y) = \frac{1}{y - z_{0}} + C\ ,
\end{equation}
and
\begin{eqnarray}
g^2(z) - 2V(z)g(z) &=& \frac{\kappa}{z - z_{0}}\ , \quad  \kappa = {2}\int_{a}^{b}\frac{f(y) \ dy}{y - z_{0}} \ ,
\nonumber\\
g(z) &=& V(z) - \sqrt{V^2(z) + \frac{\kappa}{z - z_{0}}}\ .
\end{eqnarray}
Again $g(z)$ must have a branch cut along the real line between $a$ and $b$, and this fixes the undetermined constants:
\begin{eqnarray}
C &=& \frac{1}{\sqrt{(z_{0} - a)(z_{0} - b)}}\ , \quad \kappa = -[1+C(a-z_0)]^2/(a-z_0)\ ,
\nonumber\\
g(z) &=& \frac{1}{z - z_{0}} + \frac{1}{\sqrt{(z_{0} - a)(z_{0} - b)}} - \frac{{1}}{z - z_{0}} \frac{\sqrt{(z - a)(z - b)}}{\sqrt{(z_{0} - a)(z_{0} - b)}}\ . \label{eq:a18}
\end{eqnarray}
Finally,
\begin{equation}
f(y) = \frac{1}{\pi(y - z_{0})}{\frac{\sqrt{(y - a)(b - y)}}{\sqrt{(z_{0} - a)(z_{0} - b)}}}\ .\label{eq:a19}
\end{equation}
In eqs.~(\ref{eq:a18}, \ref{eq:a19}), and in the following sections, we specify the branch $\sqrt{(z - a)(z - b)} \to z$ at large complex $z$, while $\sqrt{(y - a)(b - y)}$ is real ($y$ is restricted to the range $(a,b)$).

\subsubsection*{Special case III: $j(y) = 1/(y - z_{0})$.}   

This is simply $-\int dz_0$ of the previous source, and so linearity determines
\begin{eqnarray}
f(y) &=& \int_{z_0}^\infty \frac{dz'_0}{\pi(y - z'_{0})}{\frac{\sqrt{(y - a)(b - y)}}{\sqrt{(z'_{0} - a)(z'_{0} - b)}}} \nonumber\\
&=& \frac{i}{\pi}\ln \frac{ab + yz_0 - \frac{1}{2}(a+b)(y+z_0) + i\sqrt{(y - a)(b - y)}\sqrt{(z_{0} - a)(z_{0} - b)}}{(z_0 - y)\left(y  - \frac{1}{2}(a+b) + i \sqrt{(y - a)(b - y)}\right)}\ .\qquad \label{eq:iiisol}
\end{eqnarray}
This simplifies near the endpoints;
\begin{eqnarray}
f(y \sim b) &=& \frac{2}{\pi} \frac{\sqrt{b-y}}{\sqrt{b-a}} \Biggl( \frac{\sqrt{z_0 - a}}{\sqrt{z_0-b}} - 1
\Biggr)\ ,
\nonumber\\
f(y \sim a) &=& \frac{2}{\pi} \frac{\sqrt{y-a}}{\sqrt{b-a}} \Biggl( 1 - \frac{\sqrt{z_0 - b}}{\sqrt{z_0-a}} 
\Biggr)\ . \label{eq:iiilims}
\end{eqnarray}

\subsubsection*{Green's function solutions}

The case $j(y) = \delta(y-y_0)$, $a < y_0 < b$, is obtained from the previous solution by linearity,
\begin{equation}
\delta(y - y_{0}) = \frac{1}{2\pi i} \biggl( \frac{1}{y-y_0-i\epsilon} - \frac{1}{y-y_0+i\epsilon} \biggr)\ ,
\label{eq:poles}
\end{equation}
and so we obtain, after some rearrangement,
\begin{eqnarray}
f(y) &\equiv& h(y,y_0) 
\nonumber\\
&=&  \frac{1}{\pi^2}\ln 
\left| \frac{ab + yy_0 - \frac{1}{2}(a+b)(y+y_0) + \sqrt{(y - a)(b - y)}\sqrt{(y_0 - a)(b - y_{0})}}{(y_0 - y)(b-a)/2} \right|\ .\quad
\end{eqnarray}
This gives the solution to eq.~(\ref{eq:bulkg}) for general $j(y)$:
\begin{equation}
f(y) = \int_a^b h(y,y') j(y')\,  dy'\ . \label{eq:genj}
\end{equation}
For bounded $j$ this is the  unique bounded solution.

Let us also give a Green's function solution to the integrated equation~(\ref{eq:simp}).  A solution does not exist for all $V(x)$: for a constant $V(x)$ eq.~(\ref{eq:h2}) leads to a contradiction.
Rather, a solution exists for functions $V(x)$ satisfying one constraint.  Using eq.~(\ref{eq:poles})
for $a < y_0 < b$ one finds that for
\begin{equation}
f(y) = -\frac{1}{\pi^2(y-y_0)} \frac{\sqrt{(y-a)(b-y)}}{\sqrt{(y_0-a)(b-y_0)}}
\end{equation}
one has
\begin{equation}
\dashint_{a}^{b} \frac{f(y') \, dy'}{y - y'} =  \delta(y-y_0) - \frac{1}{\pi \sqrt{(y_0-a)(b-y_0)}}\ .
\end{equation}
Therefore, if
\begin{equation}
\int_a^b \frac{V(y_0) \, dy_0} {\sqrt{(y_0-a)(b-y_0)}} = 0 \label{eq:fhcon}
\end{equation}
then eq.~(\ref{eq:simp}) is satisifed by
\begin{equation}
f(y) = -\dashint_a^b \frac{V(y_0) }{\pi^2(y-y_0)} \frac{\sqrt{(y-a)(b-y)}}{\sqrt{(y_0-a)(b-y_0)}} dy_0\ .
\label{eq:simpgen}
\end{equation}


\begin{thebibliography}{0}
\baselineskip=14pt
\parskip = 0.025in

\bibitem{MZ}
J.~A.~Minahan and K.~Zarembo,
  ``The Bethe-ansatz for N = 4 super Yang-Mills,''
  JHEP {\bf 0303}, 013 (2003)
  [arXiv:hep-th/0212208].

\bibitem{BKS}
N.~Beisert, C.~Kristjansen and M.~Staudacher,
  ``The dilatation operator of N = 4 super Yang-Mills theory,''
  Nucl.\ Phys.\ B {\bf 664}, 131 (2003)
  [arXiv:hep-th/0303060].

\bibitem{BS1}
N.~Beisert and M.~Staudacher,
``The N = 4 SYM integrable super spin chain,''
Nucl.\ Phys.\ B {\bf 670}, 439 (2003)
[arXiv:hep-th/0307042].

\bibitem{MSW}
 G.~Mandal, N.~V.~Suryanarayana and S.~R.~Wadia,
  ``Aspects of semiclassical strings in AdS(5),''
  Phys.\ Lett.\ B {\bf 543}, 81 (2002)
  [arXiv:hep-th/0206103].
  
\bibitem{BPR} 
I.~Bena, J.~Polchinski and R.~Roiban,
  ``Hidden symmetries of the AdS(5) x S**5 superstring,''
  Phys.\ Rev.\ D {\bf 69}, 046002 (2004)
  [arXiv:hep-th/0305116].
  
\bibitem{QCD}
  L.~N.~Lipatov,
  ``High-energy asymptotics of multicolor QCD and exactly solvable lattice
  models,''
  arXiv:hep-th/9311037;\\[3pt]
  L.~D.~Faddeev and G.~P.~Korchemsky,
  ``High-energy QCD as a completely integrable model,''
  Phys.\ Lett.\ B {\bf 342}, 311 (1995)
  [arXiv:hep-th/9404173];\\[3pt]
V.~M.~Braun, S.~E.~Derkachov and A.~N.~Manashov,
  ``Integrability of three-particle evolution equations in {QCD},''
  Phys.\ Rev.\ Lett.\  {\bf 81}, 2020 (1998)
  [arXiv:hep-ph/9805225].
  
\bibitem{QCDrev}
A.~V.~Belitsky, V.~M.~Braun, A.~S.~Gorsky and G.~P.~Korchemsky,
  ``Integrability in QCD and beyond,''
  Int.\ J.\ Mod.\ Phys.\ A {\bf 19}, 4715 (2004)
  [arXiv:hep-th/0407232].

\bibitem{arkrev}
A.~A.~Tseytlin,
``Spinning strings and AdS/CFT duality,"
arXiv:hep-th/0311139.

\bibitem{Mrev}
  A.~Marshakov,
  ``Quasiclassical geometry and integrability of AdS/CFT correspondence,''
  Theor.\ Math.\ Phys.\  {\bf 142}, 222 (2005)
  [Teor.\ Mat.\ Fiz.\  {\bf 142}, 265 (2005)]
  [arXiv:hep-th/0406056].
  
\bibitem{Brev}
  N.~Beisert,
   ``The dilatation operator of N = 4 super Yang-Mills theory and
   integrability,''
  Phys.\ Rept.\  {\bf 405}, 1 (2005)
  [arXiv:hep-th/0407277].

\bibitem{Trev}
  A.~A.~Tseytlin,
   ``Semiclassical strings and AdS/CFT,''
  arXiv:hep-th/0409296.

\bibitem{Zrev}
  K.~Zarembo,
   ``Semiclassical Bethe ansatz and AdS/CFT,''
  Comptes Rendus Physique {\bf 5}, 1081 (2004)
  [Fortsch.\ Phys.\  {\bf 53}, 647 (2005)]
  [arXiv:hep-th/0411191].

\bibitem{Srev}
  I.~Swanson,
   ``Superstring holography and integrability in AdS(5) x S**5,''
  arXiv:hep-th/0505028.

\bibitem{Prev}
  J.~Plefka,
  ``Spinning strings and integrable spin chains in the AdS/CFT
  correspondence,''
  arXiv:hep-th/0507136.

\bibitem{Bstrings}
N.~Beisert, review talk at Strings 2006,
http://www.fields.utoronto.ca/audio/05-06/strings/beisert/.

\bibitem{BMN}
  D.~Berenstein, J.~M.~Maldacena and H.~Nastase,
  ``Strings in flat space and pp waves from N = 4 super Yang Mills,''
  JHEP {\bf 0204}, 013 (2002)
  [arXiv:hep-th/0202021].

\bibitem{GKP}
  S.~S.~Gubser, I.~R.~Klebanov and A.~M.~Polyakov,
  ``A semi-classical limit of the gauge/string correspondence,''
  Nucl.\ Phys.\ B {\bf 636}, 99 (2002)
  [arXiv:hep-th/0204051].
  
\bibitem{FT}
  S.~Frolov and A.~A.~Tseytlin,
  ``Multi-spin string solutions in AdS(5) x S**5,''
  Nucl.\ Phys.\ B {\bf 668}, 77 (2003)
  [arXiv:hep-th/0304255].
  
\bibitem{FT2}
  S.~Frolov and A.~A.~Tseytlin,
  ``Quantizing three-spin string solution in AdS(5) x S**5,''
  JHEP {\bf 0307}, 016 (2003)
  [arXiv:hep-th/0306130].

\bibitem{BMSZ}
  N.~Beisert, J.~A.~Minahan, M.~Staudacher and K.~Zarembo,
  ``Stringing spins and spinning strings,''
  JHEP {\bf 0309}, 010 (2003)
  [arXiv:hep-th/0306139].
  
\bibitem{BFST}
  N.~Beisert, S.~Frolov, M.~Staudacher and A.~A.~Tseytlin,
  ``Precision spectroscopy of AdS/CFT,''
  JHEP {\bf 0310}, 037 (2003)
  [arXiv:hep-th/0308117].
  
\bibitem{Kru}
  M.~Kruczenski,
  ``Spin chains and string theory,''
  Phys.\ Rev.\ Lett.\  {\bf 93}, 161602 (2004)
  [arXiv:hep-th/0311203].

\bibitem{Cetal}
  C.~G.~Callan, H.~K.~Lee, T.~McLoughlin, J.~H.~Schwarz, I.~Swanson and X.~Wu,
   ``Quantizing string theory in AdS(5) x S**5: Beyond the pp-wave,''
  Nucl.\ Phys.\ B {\bf 673}, 3 (2003)
  [arXiv:hep-th/0307032];\\[3pt]
  C.~G.~Callan, T.~McLoughlin and I.~Swanson,
   ``Holography beyond the Penrose limit,''
  Nucl.\ Phys.\ B {\bf 694}, 115 (2004)
  [arXiv:hep-th/0404007].

\bibitem{SS}
  D.~Serban and M.~Staudacher,
  ``Planar N = 4 gauge theory and the Inozemtsev long range spin chain,''
  JHEP {\bf 0406}, 001 (2004)
  [arXiv:hep-th/0401057].

\bibitem{KMMZ}
  V.~A.~Kazakov, A.~Marshakov, J.~A.~Minahan and K.~Zarembo,
  ``Classical / quantum integrability in AdS/CFT,''
  JHEP {\bf 0405}, 024 (2004)
  [arXiv:hep-th/0402207].

\bibitem{BKSZ}
  V.~A.~Kazakov and K.~Zarembo,
   ``Classical / quantum integrability in non-compact sector of AdS/CFT,''
  JHEP {\bf 0410}, 060 (2004)
  [arXiv:hep-th/0410105];\\[3pt]
  N.~Beisert, V.~A.~Kazakov and K.~Sakai,
   ``Algebraic curve for the SO(6) sector of AdS/CFT,''
  arXiv:hep-th/0410253;\\[3pt]
  S.~Schafer-Nameki,
   ``The algebraic curve of 1-loop planar N = 4 SYM,''
  Nucl.\ Phys.\ B {\bf 714}, 3 (2005)
  [arXiv:hep-th/0412254];\\[3pt]
  N.~Beisert, V.~A.~Kazakov, K.~Sakai and K.~Zarembo,
  ``The algebraic curve of classical superstrings on AdS(5) x S**5,''
  arXiv:hep-th/0502226.

\bibitem{BKSZ2}
  N.~Beisert, V.~A.~Kazakov, K.~Sakai and K.~Zarembo,
  ``Complete spectrum of long operators in N = 4 SYM at one loop,''
  JHEP {\bf 0507}, 030 (2005)
  [arXiv:hep-th/0503200].

\bibitem{BDipS}
  N.~Beisert, V.~Dippel and M.~Staudacher,
  ``A novel long range spin chain and planar N = 4 super Yang-Mills,''
  JHEP {\bf 0407}, 075 (2004)
  [arXiv:hep-th/0405001].

\bibitem{SSmat}
  M.~Staudacher,
  ``The factorized S-matrix of CFT/AdS,''
  JHEP {\bf 0505}, 054 (2005)
  [arXiv:hep-th/0412188].
  
\bibitem{BS05}
  N.~Beisert and M.~Staudacher,
  ``Long-range PSU(2,2$|$4) Bethe ansaetze for gauge theory and strings,''
  arXiv:hep-th/0504190.

\bibitem{Vallilo}
  B.~C.~Vallilo,
   ``Flat currents in the classical AdS(5) x S**5 pure spinor superstring,''
  JHEP {\bf 0403}, 037 (2004)
  [arXiv:hep-th/0307018].
  
\bibitem{Berk}
  N.~Berkovits,
   ``BRST cohomology and nonlocal conserved charges,''
  JHEP {\bf 0502}, 060 (2005)
  [arXiv:hep-th/0409159];
   ``Quantum consistency of the superstring in AdS(5) x S**5 background,''
  JHEP {\bf 0503}, 041 (2005)
  [arXiv:hep-th/0411170].

\bibitem{AFS}
  G.~Arutyunov, S.~Frolov and M.~Staudacher,
  ``Bethe ansatz for quantum strings,''
  JHEP {\bf 0410}, 016 (2004)
  [arXiv:hep-th/0406256].
  
\bibitem{Bqu}
  N.~Beisert,
  ``Spin chain for quantum strings,''
  Fortsch.\ Phys.\  {\bf 53}, 852 (2005)
  [arXiv:hep-th/0409054].
  
\bibitem{oneloop}
  S.~A.~Frolov, I.~Y.~Park and A.~A.~Tseytlin,
   ``On one-loop correction to energy of spinning strings in S(5),''
  Phys.\ Rev.\ D {\bf 71}, 026006 (2005)
  [arXiv:hep-th/0408187];\\[3pt]
  I.~Y.~Park, A.~Tirziu and A.~A.~Tseytlin,
   ``Spinning strings in AdS(5) x S**5: One-loop correction to energy in  SL(2)
   sector,''
  JHEP {\bf 0503}, 013 (2005)
  [arXiv:hep-th/0501203];\\[3pt]
    N.~Beisert, A.~A.~Tseytlin and K.~Zarembo,
   ``Matching quantum strings to quantum spins: One-loop vs. finite-size
   corrections,''
  Nucl.\ Phys.\ B {\bf 715}, 190 (2005)
  [arXiv:hep-th/0502173];\\[3pt]
  R.~Hernandez, E.~Lopez, A.~Perianez and G.~Sierra,
   ``Finite size effects in ferromagnetic spin chains and quantum  corrections
   to classical strings,''
  JHEP {\bf 0506}, 011 (2005)
  [arXiv:hep-th/0502188];\\[3pt]
  H.~Fuji and Y.~Satoh,
   ``Quantum fluctuations of rotating strings in AdS(5) x S**5,''
  arXiv:hep-th/0504123;\\[3pt]
   S.~Schafer-Nameki, M.~Zamaklar and K.~Zarembo,
  ``Quantum corrections to spinning strings in AdS(5) x S**5 and Bethe ansatz:
  A comparative study,''
  arXiv:hep-th/0507189.

\bibitem{charges}
L.~Dolan, C.~R.~Nappi and E.~Witten,
  ``A relation between approaches to integrability in superconformal
  Yang-Mills theory,''
  JHEP {\bf 0310}, 017 (2003)
  [arXiv:hep-th/0308089];\\[3pt]
A.~Gorsky,
  ``Spin chains and gauge / string duality,''
  Theor.\ Math.\ Phys.\  {\bf 142}, 153 (2005)
  [Teor.\ Mat.\ Fiz.\  {\bf 142}, 179 (2005)]
  [arXiv:hep-th/0308182];\\[3pt]
L.~F.~Alday,
  ``Non-local charges on AdS(5) x S**5 and pp-waves,''
  JHEP {\bf 0312}, 033 (2003)
  [arXiv:hep-th/0310146];\\[3pt]
G.~Arutyunov and M.~Staudacher,
  ``Matching higher conserved charges for strings and spins,''
  JHEP {\bf 0403}, 004 (2004)
  [arXiv:hep-th/0310182];\\[3pt]
 J.~Engquist,
  ``Higher conserved charges and integrability for spinning strings in  AdS(5)
  x S**5,''
  JHEP {\bf 0404}, 002 (2004)
  [arXiv:hep-th/0402092];\\[3pt]
A.~M.~Polyakov,
  ``Conformal fixed points of unidentified gauge theories,''
  Mod.\ Phys.\ Lett.\ A {\bf 19}, 1649 (2004)
  [arXiv:hep-th/0405106];\\[3pt]
A.~Agarwal and S.~G.~Rajeev,
  ``The dilatation operator of N = 4 SYM and classical limits of spin chains
  and matrix models,''
  Mod.\ Phys.\ Lett.\ A {\bf 19}, 2549 (2004)
  [arXiv:hep-th/0405116];  
  ``Yangian symmetries of matrix models and spin chains: The dilatation
  operator of N = 4 SYM,''
  arXiv:hep-th/0409180;\\[3pt]
B.~Y.~Hou, D.~T.~Peng, C.~H.~Xiong and R.~H.~Yue,
  ``The affine hidden symmetry and integrability of type IIB superstring in
  AdS(5) x S**5,''
  arXiv:hep-th/0406239;\\[3pt]
M.~Hatsuda and K.~Yoshida,
  ``Classical integrability and super Yangian of superstring on AdS(5) x
  S**5,''
  arXiv:hep-th/0407044;\\[3pt]
A.~Mikhailov,
  ``Notes on fast moving strings,''
  arXiv:hep-th/0409040;
  ``Anomalous dimension and local charges,''
  arXiv:hep-th/0411178; 
  ``Plane wave limit of local conserved charges,''
  arXiv:hep-th/0502097;
    ``Baecklund transformations, energy shift and the plane wave limit,''
  arXiv:hep-th/0507261;\\[3pt]
I.~Swanson,
  ``Quantum string integrability and AdS/CFT,''
  Nucl.\ Phys.\ B {\bf 709}, 443 (2005)
  [arXiv:hep-th/0410282];\\[3pt]
  G.~Arutyunov and S.~Frolov,
  ``Integrable Hamiltonian for classical strings on AdS(5) x S**5,''
  JHEP {\bf 0502}, 059 (2005)
  [arXiv:hep-th/0411089].;\\[3pt]
  A.~Das, J.~Maharana, A.~Melikyan and M.~Sato,
  ``The algebra of transition matrices for the AdS(5) x S**5 superstring,''
  JHEP {\bf 0412}, 055 (2004)
  [arXiv:hep-th/0411200];\\[3pt]
L.~F.~Alday, G.~Arutyunov and A.~A.~Tseytlin,
  ``On integrability of classical superstrings in AdS(5) x S**5,''
  JHEP {\bf 0507}, 002 (2005)
  [arXiv:hep-th/0502240];\\[3pt]
C.~A.~S.~Young,
  ``Non-local charges, Z(m) gradings and coset space actions,''
  arXiv:hep-th/0503008\\[3pt]
B.~Chen, Y.~L.~He, P.~Zhang and X.~C.~Song,
  ``Flat currents of the Green-Schwarz superstrings in AdS(5) x S**1 and AdS(3)
  x S**3 backgrounds,''
  Phys.\ Rev.\ D {\bf 71}, 086007 (2005)
  [arXiv:hep-th/0503089];\\[3pt]
G.~Arutyunov and M.~Zamaklar,
  ``Linking Baecklund and monodromy charges for strings on AdS(5) x S**5,''
  JHEP {\bf 0507}, 026 (2005)
  [arXiv:hep-th/0504144];\\[3pt]
N.~Drukker and B.~Fiol,
  ``On the integrability of Wilson loops in AdS(5) x S**5: Some periodic
  ansatze,''
  arXiv:hep-th/0506058;\\[3pt]
E.~Abdalla and A.~Lima-Santos,
  ``Classical non-local conserved charges in string theory,''
  arXiv:hep-th/0506148;\\[3pt]
A.~Das, A.~Melikyan and M.~Sato,
  ``The algebra of flat currents for the string on $AdS_5 x S^5$ in the
  light-cone gauge,''
  arXiv:hep-th/0508183.

\bibitem{MP}
 N.~Mann and J.~Polchinski,
  ``Finite density states in integrable conformal field theories,''
  in {\it From Fields to Strings: Circumnavigating Theoretical Physics, in memory of Ian Kogan,} ed. M. Shifman et al. (World Scientific, Singapore, 2005)
[arXiv:hep-th/0408162].
  
\bibitem{SWK}
  H.~Saleur and B.~Wehefritz-Kaufmann,
  ``Integrable quantum field theories with OSP(m/2n) symmetries,''
  Nucl.\ Phys.\ B {\bf 628}, 407 (2002)
  [arXiv:hep-th/0112095].
  
\bibitem{ZZS}
A.~B.~Zamolodchikov and A.~B.~Zamolodchikov,
  ``Factorized S-matrices in two dimensions as the exact solutions of certain
  relativistic quantum field models,''
  Annals Phys.\  {\bf 120}, 253 (1979).
  
\bibitem{BVW}
N.~Berkovits, C.~Vafa and E.~Witten,
  ``Conformal field theory of AdS background with Ramond-Ramond flux,''
  JHEP {\bf 9903}, 018 (1999)
  [arXiv:hep-th/9902098].
  
\bibitem{BSZ}
M.~Bershadsky, S.~Zhukov and A.~Vaintrob,
  ``PSL(n$|$n) sigma model as a conformal field theory,''
  Nucl.\ Phys.\ B {\bf 559}, 205 (1999)
  [arXiv:hep-th/9902180].
  
\bibitem{zirn}
M.~R.~Zirnbauer,
  ``Conformal field theory of the integer quantum Hall plateau transition,''
  arXiv:hep-th/9905054.

\bibitem{ParSour}
  G.~Parisi and N.~Sourlas,
   ``Random magnetic fields, supersymmetry and negative dimensions,''
  Phys.\ Rev.\ Lett.\  {\bf 43}, 744 (1979).
  
\bibitem{Weg}
  F.~Wegner,
   ``Four loop order beta function of nonlinear sigma models in symmetric
   spaces,''
  Nucl.\ Phys.\ B {\bf 316}, 663 (1989).

\bibitem{LPohl}
M.~Luscher and K.~Pohlmeyer,
   ``Scattering of massless lumps and nonlocal charges in the two-dimensional
   classical nonlinear sigma model,''
  Nucl.\ Phys.\ B {\bf 137}, 46 (1978).
   
\bibitem{BIZZ}  
   E.~Brezin, C.~Itzykson, J.~Zinn-Justin and J.~B.~Zuber,
  ``Remarks about the existence of nonlocal charges in two-dimensional
  models,''
  Phys.\ Lett.\ B {\bf 82}, 442 (1979).

\bibitem{ZZ1}
A.~B.~Zamolodchikov and A.~B.~Zamolodchikov,
 ``Massless factorized scattering and sigma models with topological terms,''
{\it Nucl.\ Phys.}\ {\bf B379}, 602 (1992).

\bibitem{FS}
P.~Fendley and H.~Saleur,
   ``Massless integrable quantum field theories and massless scattering in
   (1+1)-dimensions,''
[arXiv:hep-th/9310058].

\bibitem{bethe}
H.~Bethe,
  ``On the theory of metals. 1. Eigenvalues and eigenfunctions for the linear
  atomic chain,''
  Z.\ Phys.\  {\bf 71}, 205 (1931).

\bibitem{thacker}
H.~B.~Thacker,
  ``Exact integrability in quantum field theory and statistical systems,''
  Rev.\ Mod.\ Phys.\  {\bf 53}, 253 (1981).

\bibitem{yy}
C.~N.~Yang and C.~P.~Yang,
  ``One-dimensional chain of anisotropic spin spininteractions. 1. Proof of
  Bethe's hypothesis for ground state in a finite system,''
  Phys.\ Rev.\  {\bf 150}, 321 (1966).
  

\bibitem{RS}
N.~Read and H.~Saleur,
``Exact spectra of conformal supersymmetric nonlinear sigma models in two dimensions,''
{\it Nucl.\ Phys.}\ {\bf B613}, 409 (2001)
[arXiv:hep-th/0106124].

\bibitem{JNW}
G.~I.~Japaridze, A.~A.~Nersesian and P.~B.~Wiegmann,
  ``Exact results in the two-dimensional U(1) symmetric thirring model,''
  Nucl.\ Phys.\ B {\bf 230}, 511 (1984).
  
\bibitem{has2}
P.~Hasenfratz and F.~Niedermayer,
  ``The exact mass gap of the O(N) sigma model for arbitrary N $\geq 3$ in D   2,''
  Phys.\ Lett.\ B {\bf 245}, 529 (1990).
  
\bibitem{yang}
C.~N.~Yang,
  ``Some exact results for the many body problems in one dimension with
 repulsive delta function interaction,''
  Phys.\ Rev.\ Lett.\  {\bf 19}, 1312 (1967).

\bibitem{DfGZj}
  P.~Di Francesco, P.~H.~Ginsparg and J.~Zinn-Justin,
  ``2-D Gravity and random matrices,''
  Phys.\ Rept.\  {\bf 254}, 1 (1995)
  [arXiv:hep-th/9306153].

\bibitem{PW}
A.~M.~Polyakov and P.~B.~Wiegmann,
  ``Theory Of Nonabelian Goldstone Bosons In Two Dimensions,''
  Phys.\ Lett.\ B {\bf 131}, 121 (1983).
  
\bibitem{Wieg}
 P.~Wiegmann,
  ``Exact Factorized S Matrix Of The Chiral Field In Two-Dimensions,''
  Phys.\ Lett.\ B {\bf 142}, 173 (1984).

\bibitem{FadR}
 L.~D.~Faddeev and N.~Y.~Reshetikhin,
  ``Integrability Of The Principal Chiral Field Model In (1+1)-Dimension,''
  Annals Phys.\  {\bf 167}, 227 (1986).


\end{thebibliography}
\end{document}